\documentclass[final,5p,times,twocolumn]{elsarticle}

\usepackage{amssymb}
\usepackage{graphicx, subcaption}
\usepackage{amsmath}

\journal{ Applications in Energy and Combustion Science }
\pdfoutput=1

\begin{document}

\begin{frontmatter}



\title{The Backward Problem in Plasma-Assisted Combustion: Experiments of Nanosecond Pulsed Discharges Driven by Flames}


\author[inst1]{Carmen Guerra-Garcia}
\affiliation[inst1]{organization={Department of Aeronautics and Astronautics},
            addressline={77 Massachusetts Avenue}, 
            city={Cambridge},
            postcode={02139}, 
            state={MA},
            country={USA}}

\author[inst1]{Colin A. Pavan}

\begin{abstract}
Plasma technologies are a promising way of addressing a number of challenges in combustion, ranging from stabilization and flame-holding for hypersonic vehicles, to enabling low-emissions propulsion and power. Whereas most works in the literature have centered on the forward problem: the impact of the plasma on different combustion metrics, including static and dynamic flame stabilization; fewer studies have been devoted to the backward problem: or the significance of having a strongly inhomogeneous environment, often unsteady in time, on the discharge characteristics. In this perspectives article, the backward problem is addressed  in a hierarchical manner. First, the implications of the discharge proceeding in an inhomogeneous background gas, with gradients in density and composition, and nonuniform flows, are discussed. Next, these foundation experiments are used as elementary building blocks to explain experiments of nanosecond pulsed discharges in dynamic combustion environments, including a transient laminar flame and a combustor undergoing dynamics. Finally, the strong coupling between the forward and backward problem is illustrated with an imaging study, using transparent electrodes, of a mesoscale premixed flame and a nanosecond pulsed dielectric barrier discharge. The experiments demonstrate the importance of considering the pulse-to-pulse evolution due to changes in the background environment when using plasma assistance in unsteady combustion processes, including active control of combustion dynamics.

\end{abstract}




\end{frontmatter}


\section{Introduction}
\label{sec:intro}
\subsection{Challenges in Combustion that can Benefit from Plasma-Assistance}

Plasma technologies offer an extra knob to enhance and control combustion regimes that are inherently challenging. For aerospace applications, these challenging conditions appear as a result of extreme environments intrinsic to the field itself, e.g., supersonic and hypersonic flight; or driven by the need to address environmental concerns. 

For hypersonic flight, reliable thrust and high propulsion efficiency over a wide range of flight Mach numbers, typically 5-25, and altitudes, 20-55km, is needed \cite{Heiser1994}. This demands effective strategies for combustion stabilization and flameholding in an extremely hostile environment, with comparable timescales of the fluid flow and the chemical kinetics. Viable methods to achieve stabilization include (i) mixing enhancement of the fuel and air streams, by introducing turbulence and accelerating molecular diffusion; and (ii) chemistry enhancement, by accelerating the chain reaction process through either a local temperature increase or by generating a pool of radicals and excited species. For different conditions, pursuing either chemistry or mixing enhancement can be more effective. E.g., whereas pure mixing methods, such as vortex generators, can be efficient in the lower Mach range; incorporating chemical enhancement, such as heat recirculation using cavity flame holders, is beneficial at higher Mach \cite{Liu2020}. In the case of plasma, all three effects can be accessed: heat, reactive radicals and excited species, as well as turbulent kinetic energy and vorticity generation \cite{Gann2015,Leonov2009}. 

From the environmental perspective, transportation contributes to climate change and global warming through a complex set of processes including CO$_2$ emissions, NO$_\text{x}$ emissions, water vapor, soot, and sulfate aerosols. In particular, air transportation contributes over 2\% of all man-made carbon dioxide, but its impact on the effective radiative forcing is about three times higher due to contributions from nitrogen oxides and contrail cirrus \cite{Lee2020The2018}. The general approach to reduce NO$_\text{x}$ emissions is to reduce the process temperature, by burning lean and ultra-lean mixtures \cite{Turns2000}, but these mixtures are prone to combustion dynamics, that can cause wear and damage to combustor components and in extreme cases downstream turbine elements \cite{Lieuwen2012}. Hence, novel technologies are needed to extend the static stability (lean blowoff) limits of flames, as well as provide a means for suppressing combustion dynamics. To address the carbon emissions, sustainable fuels will be introduced. In the near term, these will be drop-in fuels with similar physical and chemical properties to conventional fuels but about 80\% lower CO$_2$ emissions on a lifecycle basis, and therefore similar combustion challenges to be addressed in addition to fuel-flexibility constraints. In the longer term, the propulsion system will need to be significantly modified to safely burn hydrogen, or reliably and stably burn its carrier ammonia \cite{Masri2021,Kobayashi2019ScienceCombustion}. Plasma technologies for ignition and flame stabilization may mitigate many of the challenges associated with using these novel fuels in propulsion.

\subsection{The Forward Problem: Plasma on Flame}
A broad classification of the literature of plasma-assisted combustion can be made in terms of demonstration works, that testify to the ability of plasma to promote ignition and flame stability; and fundamental works, that clarify the energy pathways activated by the plasma that enhance combustion.  

From the practical perspective, numerous demonstrations for hydrocarbon fuels and hydrogen are reported in the literature including: reductions in ignition delay times and minimum ignition temperatures, flame speed enhancement, extensions of flammability limits, and mitigation of combustion dynamics \cite{Ju2015PlasmaChemistry,Starikovsky2015,Adamovich2015ChallengesCombustion,Shinohara2009,Stockman2009,Yu2010,Elkholy2018BurningPlatform,Ombrello2010_1}. Related to the emissions concern, there is compelling evidence that plasma can extend the lean blowout limit \cite{Pilla2006,Pham2011StabilizationDischarges,DiSabatino2020EnhancementDischarges}, with Barbosa et al. reporting a three-fold extension in the equivalence ratio for extinction \cite{Barbosa2015InfluenceCombustor}. It is still unclear however whether this will have significant impact on the NO$_\text{x}$ emissions, as the plasma activates new pathways of NO$_\text{x}$ production \cite{Choe2018BlowoffFlow}. Similar strategies have also been successful in mitigating dynamic instabilities, including suppression of pressure oscillations \cite{Pilla2006,Pham2011StabilizationDischarges,DiSabatino2020EnhancementDischarges,Barbosa2015InfluenceCombustor,Moeck2013}, with recent work by Shanbhogue et al. \cite{Shanbhogue2022} demonstrating significant attenuation of large (a few percent of the operating condition) amplitude pressure oscillations in a swirl-stabilized burner. Current efforts have also addressed plasma-enhancement of ammonia flames \cite{Faingold2021APlasma,Shioyoke2018NumericalFlame,Taneja2021}, particularly exciting are the results by Choe et al. \cite{Choe2021PlasmaEnhancement} showing 35\% lean limit extension, in this case accompanied by significant NO$_\text{x}$ emissions reduction. 

\begin{figure}[htb]
    \centering
\includegraphics[width=0.45\textwidth, trim={1cm 16.0cm 2cm 1.2cm}, clip]{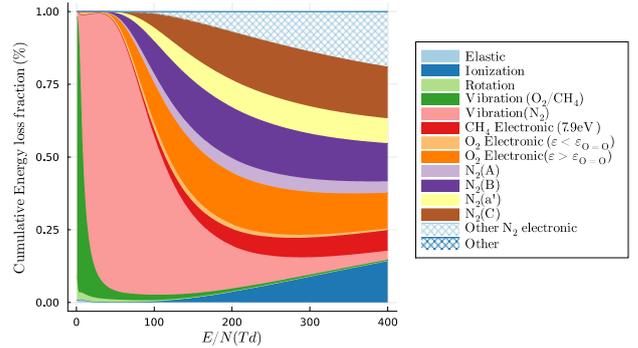}
    \caption{Fractional energy deposition into different internal energy modes for a stoichiometric methane-air mixture, as a function of the reduced electric field. Calculated using BOLSIG+\cite{Hagelaar2005} using cross sections from \cite{lxcat_hayashi,lxcat_phelps,lxcat_siglo}. The area of each region at a given value of E/N represents the relative contribution of that loss mechanism; for example, at 300Td about 5\% of the total input energy is consumed to produce N$_2$(A).}
    \label{fig:energy_share_EN}
\end{figure}

At a fundamental level, the pathways responsible for these combustion enhancements include: kinetic, thermal, and transport effects \cite{Ju2015PlasmaChemistry,Starikovsky2015}. Kinetic effects refer to ionization, excitation, and dissociation triggered by electron-impact reactions. Thermal effects refer to local heating of the gas that in turn increases the Arrhenius-type reaction rates, and can follow kinetic effects through slow vibrational relaxation processes, in microsecond-timescales, and fast electronic relaxation processes, in nanosecond-timescales \cite{Popov2022RelaxationStabilization}. Transport effects include ion-driven winds \cite{Guerra-Garcia2015}, modified diffusivities, and pressure wave generation \cite{Xu2011}. Specific to plasma-strategies are the kinetic effects, namely electron-impact reactions whose reaction rates are dependent on the electron temperature rather than the significantly lower temperature of the gas. How the electrical energy is expended can therefore be controlled by the reduced electric field (electric field divided by neutral gas number density, or $E/N$\footnote{The reduced electric field is measured in townsend units: 1 townsend, or 1 Td, is
equivalent to 250 V applied across a 1 cm gap at standard pressure and temperature.}) as the electrons gain energy from the electric field and spend that energy in collisions \cite{Raizer1991,Starikovskiy2013,Nagaraja2013Multi-scaleGeometry}. At reduced fields beyond about 150 Td (easily achieved using pulsed discharges of tens of nanoseconds duration), a significant energy deposition pathway becomes electronic excitation of nitrogen and oxygen, figure~\ref{fig:energy_share_EN}. The quenching of these excited species leads to: (1) radical seeding, including oxygen radicals, providing a useful method of bypassing chain initiation reactions \cite{Popov2007,Aleksandrov2010,Starikovskiy2012Plasma-assistedTransition}, and (2) fast gas heating \cite{Popov2007,Popov2022RelaxationStabilization}, that occurs at nanosecond time scales. A recent review by Popov and Starikovskaia \cite{Popov2022RelaxationStabilization} provides an overview of the kinetics of fast gas heating, and considers the dependency of the fraction of energy spent on this process as a function of $E/N$. 

The absolute concentration of active species produced by the plasma can be further determined by the energy density deposited in the gas, or equivalently the electron number density. E.g., the concentration of atomic oxygen produced through electron impact dissociation of molecular oxygen, ($e^- + \text{O}_2 \rightarrow e^- + \text{O} + \text{O}$), can be estimated by an energy balance as:

\begin{equation}
[\text{O}]=\frac{\varepsilon_{tot}\cdot f_{\text{O}_2(dis)}\cdot N_A}{\varepsilon_{\text{O}=\text{O}}/2},
\end{equation}

where $\varepsilon_{tot}$ is the total energy density of the discharge in $J/m^3$, $f_{\text{O}_2(dis)}$ is the fraction of energy that goes to dissociation of molecular oxygen and depends on $E/N$ (figure~\ref{fig:energy_share_EN}), $\varepsilon_{\text{O}=\text{O}}$ is the $\text{O}=\text{O}$ bond energy in $J/mol$, and $N_A$ is the Avogadro number. With the cross-sections used to produce figure~\ref{fig:energy_share_EN}, the electron impact dissociation of oxygen is a two-step process caused by rapid quenching of electronically excited O$_2$ produced by the pathway labelled ``O$_2$ Electronic ($\varepsilon>\varepsilon_{\text{O}=\text{O}} $)''. The fraction of energy used for direct dissociation of O$_2$, $f_{\text{O}_2(dis)}$, is approximately given by this value.\footnote{When the electronically excited molecule of O$_2$ undergoes dissociative quenching, additional energy beyond the threshold $\varepsilon_{\text{O}=\text{O}}$ is converted to kinetic or internal energy in the product species/third body quencher and so the energy actually expended on O$_2$ dissociation will be less than the energy used to electronically excite O$_2$ to an internal energy state with energy $>\varepsilon_{\text{O}=\text{O}} $.} Similarly, the fraction of energy going into methane dissociation could be estimated using the pathway labelled ``CH$_4$ Electronic (7.9eV)'' and assuming these electronically excited states rapidly underwent dissociative quenching. This is for direct dissociation only; additional dissociation of O$_2$ and CH$_4$ will be caused when electronically excited states of nitrogen quench by collision with these molecules. At a fundamental level then, $E/N$ and $\varepsilon_{tot}$, or equivalently electron temperature and electron density, determine the plasma-activated chemistry. 

\subsection{The Backward Problem: Flame on Plasma}
Based on the above description, it is tempting to envision being able to externally control the two relevant fundamental parameters, $E/N$ and $\varepsilon_{tot}$, to favor certain reaction pathways and deliver specific amounts of plasma-activated species to affect combustion.  Although some level of control is indeed possible by adequately choosing the plasma source, electrical parameters, positioning and timing, these parameters tend to self-adjust as the plasma is generated, e.g., for applied fields above the breakdown threshold ($E/N\sim$ 120Td in air) the electron concentration rapidly grows effectively shielding the field. Moreover, a flame is an unusual environment for a gas discharge to proceed. Whereas electrical breakdown is typically studied in homogeneous and uniform gas conditions (constant composition, pressure, and temperature), and the main source of inhomogeneity comes from the background electric field (e.g., for needle electrodes) \cite{Dhali1987,Hagelaar2000,Pancheshnyi2008}; when using plasmas for combustion applications, these assumptions no longer hold. Combustion processes inherently constitute inhomogeneous environments (flame fronts, hot spots, concentration gradients, etc.), and this background gas will guide the accessible $E/N$ and $\varepsilon_{tot}$ values, the plasma regimes encountered, the coupling of the electrical energy, and even feed back into the combustion response that can be achieved. In addition, most combustion processes of practical interest present turbulent conditions, and/or transient behaviour, requiring a consideration of the dynamic environment. A solid understanding of how the combustion background affects the plasma, in static and dynamic conditions, is therefore crucial to ensure the control of the plasma source and enable the design of predictable and robust flame modification and actuation strategies. 

\subsection{Scope of this Work}
As seen in the previous sections, plasma can be used to modify and control flames, and this is the practical application the community is interested in. However, the combustion environment equally affects the plasma dynamics and kinetics, limiting our ability to control the energy deposition process. Of course in reality the two-way problem of plasma-assisted combustion is fully coupled and the effects of plasma-on-combustion and combustion-on-plasma need to be simultaneously considered. Whereas the majority of the literature focuses on the forward problem, Lacoste \cite{Lacoste2022} identified in her recent review that the effects of pressure, mixture composition and flow on the plasma, at relevant engine conditions, are relatively poorly documented. This article aims to be a first step at filling that literature gap. The scope of this work is to present the authors' perspectives on the backward problem. The focus is on the plasma dynamics of Nanosecond Pulsed Discharges (NPD) in subsonic flows, gaseous blends, laminar and turbulent flames, in premixed and nonpremixed conditions. 

This article is organized as follows. Section~\ref{sec:nonuniform} describes, from a gas discharge physics perspective, the relevant combustion processes that will serve as the baseline environments for the studies in the next sections. Sections~\ref{sec:nonuniform steady} and~\ref{sec:dynamic} summarize prior work by the authors and collaborators documenting the development of NPD plasmas in steady, but nonuniform, flame environments and their elementary surrogates; as well as dynamic combustion environments, both laminar and turbulent. Section~\ref{sec:new_results} extends these works through an imaging study of the response of a nanosecond pulsed dielectric barrier discharge to the passage of a transient laminar premixed flame. The images present a visualization of the discharge-flame interaction not observed before, enabled by the use of transparent electrodes. Section~\ref{sec:applications} summarizes the implications of the results presented on the forward problem. Finally, section~\ref{sec:conclusions} presents the conclusions of this work.

\section{Combustion Environments}
\label{sec:nonuniform}

The applicability of plasmas to enhance combustion processes has been demonstrated across various combustion modes including supersonic combustion, homogeneous ignition, premixed flames, and nonpremixed flames. In all of these environments, the gas state (composition, temperature, and pressure) and velocity field will be varying, either as a function of time (for homogeneous processes), space (for the case of localized combustion zones or flames), or both (for unsteady flames). These inhomogeneous, unsteady, backgrounds allow for two different scenarios that need to be accounted for, when designing a plasma-actuation strategy. 

The first scenario happens when the discharge gap is comparable to, or greater than, the characteristic length scale for appreciable changes in temperature, pressure, composition or velocity. Examples are given in section~\ref{sec:academic} and figure~\ref{fig:academic_burners}. In this situation, the discharge will develop in a highly inhomogeneous environment, leading to inhomogeneous plasma generation and energy deposition, and possible reinforcement of the plasma in regions where ionization is favored (e.g., hot spots). 

The second scenario happens when the discharge gap is smaller than the characteristic length scale for appreciable changes in temperature, pressure, composition or velocity. In this case, the plasma source can be contained in a locally uniform region, and examples are given in section~\ref{sec:industrial} and figure~\ref{fig:practical_burners}. In this case, if the combustion process is in steady state, the question is where to locate the plasma to maximize the benefits of plasma-assistance, considering also that the plasma characteristics will depend on this choice. If the combustion process is unsteady, e.g., a flame undergoing thermoacoustic oscillations, it is to be expected that the plasma state will be evolving at the timescale of combustion dynamics, with implications on the response that can be achieved. In what follows, some relevant combustion environments will be described from the electrical breakdown perspective. 

\subsection{Laminar Premixed and Nonpremixed Flames}\label{sec:academic}

\begin{figure}[htb]
     \centering
     \begin{subfigure}[b]{0.45\textwidth}
         \centering
         \includegraphics[width=1\textwidth, trim={0cm 0cm 0cm 0cm}, clip]{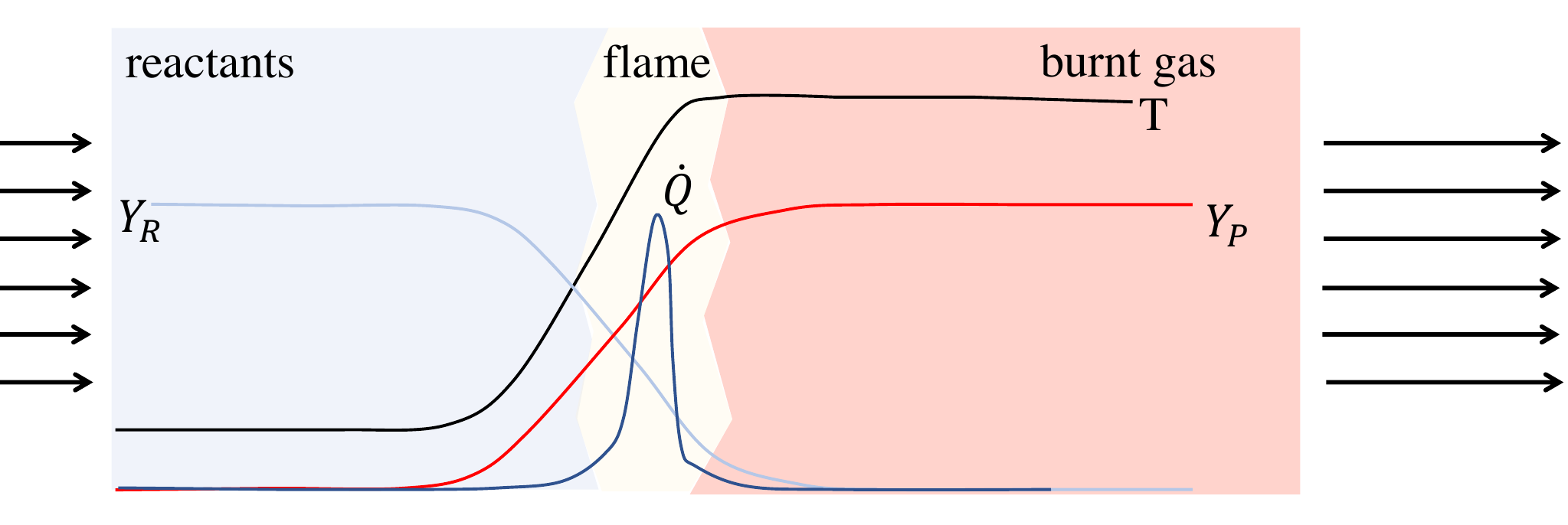}
         \caption{Laminar premixed flame.}
         \label{fig:premixed}
     \end{subfigure}
     \hfill
     \begin{subfigure}[b]{0.45\textwidth}
         \centering
         \includegraphics[width=1\textwidth, trim={0cm 0cm 0cm 0cm}, clip]{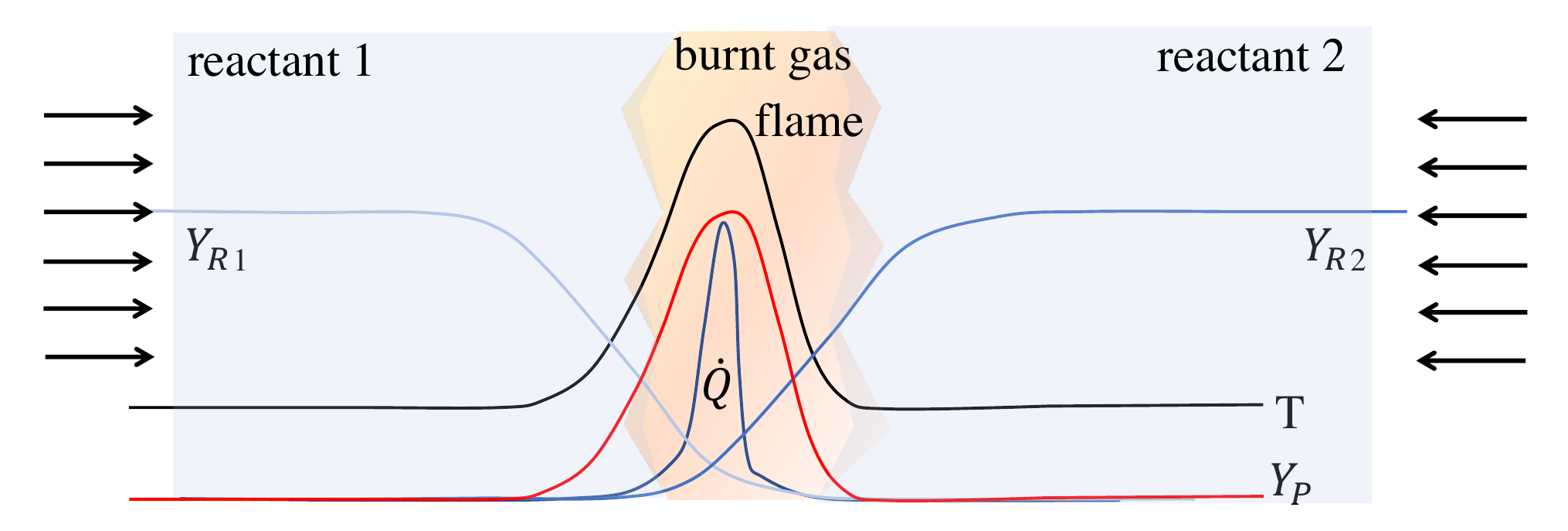}
         \caption{Laminar nonpremixed flame.}
         \label{fig:diffusion}
     \end{subfigure}
        \caption{Inhomogeneous conditions for 1D premixed and nonpremixed laminar flames. These simple environments serve as building blocks for more complex combustion processes, e.g., turbulent flames can be pictured as ensembles of laminar flame elements (or flamelets) embedded in a turbulent flow field.}
        \label{fig:academic_burners}
\end{figure}

The first classification of flames can be made in terms of whether the reactants are premixed or not. Premixed flames require that the fuel and oxidizer form a solution, and are characterized by the self-sustaining propagation of a localized combustion zone or discrete combustion wave, figure~\ref{fig:premixed}. Nonpremixed flames form when the fuel and oxidizer are originally separated, they have no characteristic speed associated to them, and typically burn more slowly than premixed flames, figure~\ref{fig:diffusion}. In both cases, the characteristic length scale for changes in the state of the gas is of the order of millimeters (at atmospheric pressure), e.g., the temperature varies by a factor of 4-7 within the thin thermal mixing layer of the laminar nonpremixed flame depicted in figure~\ref{fig:diffusion}.

In the plasma-assisted combustion literature, academic premixed and nonpremixed burners have been used to isolate the effect of different plasma species \cite{Ombrello2010_1}; to explore different actuation strategies, including pre-conditioning of the reactants (by placing the plasma ahead of the flame) \cite{Sun2011_2} and in-situ plasma deposition (by overlapping plasma and flame) \cite{Sun2013}; in addition to exploring the impact of plasma on different global combustion parameters, including the flame speed \cite{Stange2005}.

\subsection{Bluff Body- and Swirl-Stabilized Flames}\label{sec:industrial}

\begin{figure}[htb]
     \centering
     \begin{subfigure}[b]{0.45\textwidth}
         \centering
         \includegraphics[height=0.6\textwidth, trim={0cm 0cm 0cm 0cm}, clip]{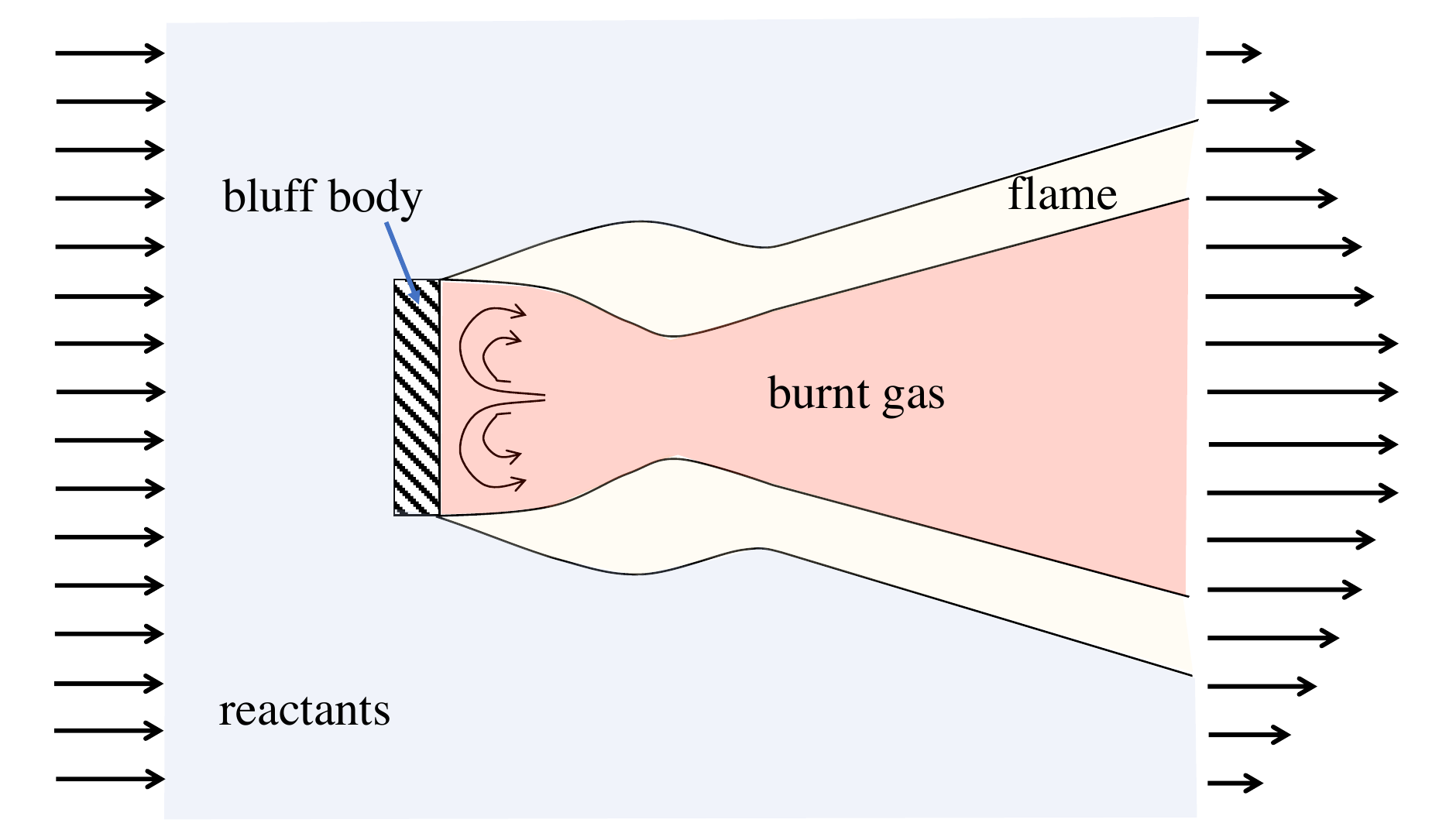}
         \caption{Bluff body-stabilized premixed flame.}
         \label{fig:bluff}
     \end{subfigure}
     \hfill
     \begin{subfigure}[b]{0.45\textwidth}
         \centering
         \includegraphics[height=0.6\textwidth, trim={0cm 0cm 0cm 0cm}, clip]{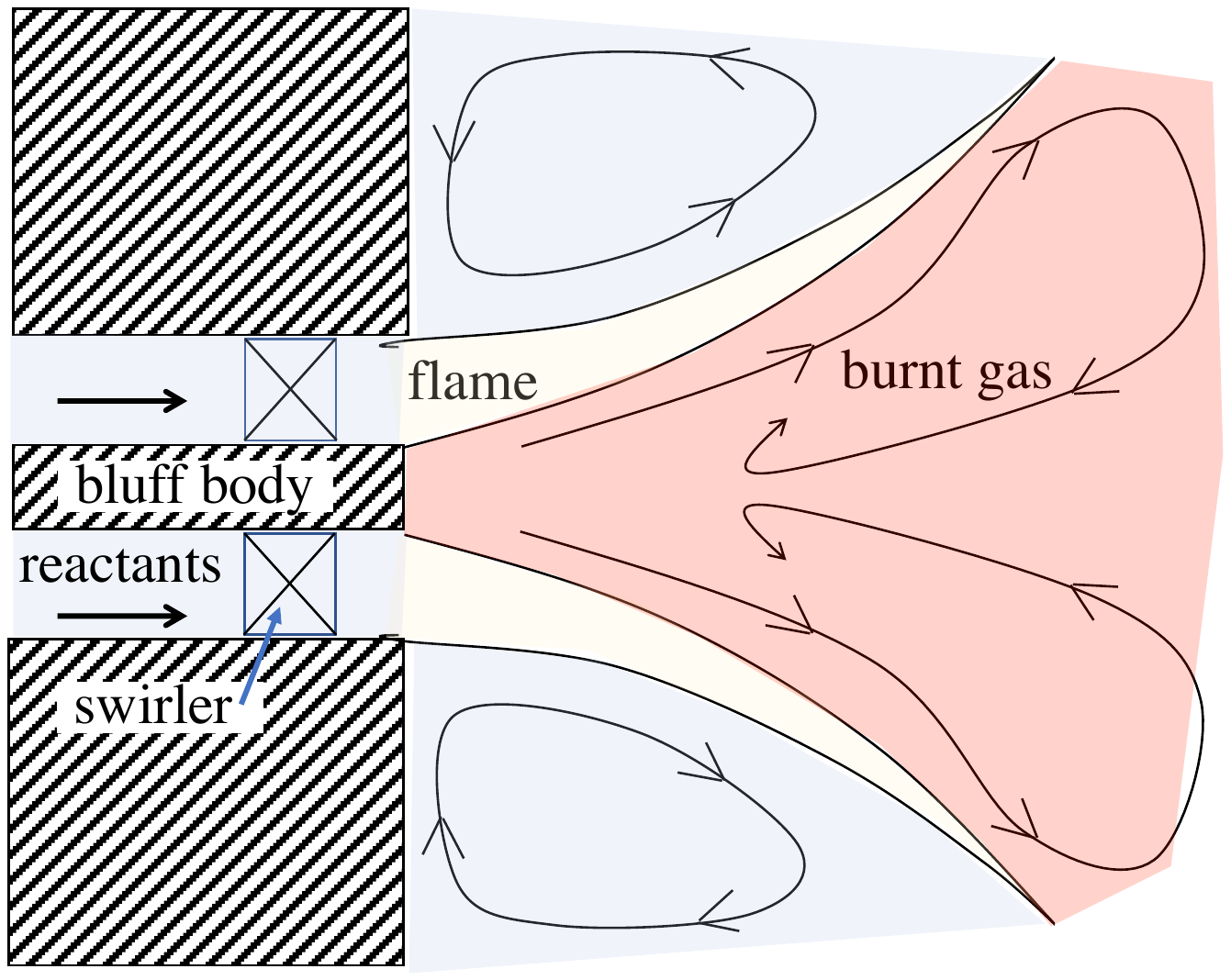}
         \caption{Swirl-stabilized premixed flame.}
         \label{fig:swirl}
     \end{subfigure}
        \caption{Inhomogeneous gas background for practical flame stabilization systems, including flow recirculation introduced by bluff bodies and swirlers.}
        \label{fig:practical_burners}
\end{figure}

Since the propagation speed of flames is much lower than local flow speeds in any practical system, practical burners typically require some flame stabilization scheme to anchor the flame. These can be realized using geometric alterations, including rapid expansions or bluff bodies, as well as swirlers, to locally slow down the flow as well as recirculate hot products, figure~\ref{fig:practical_burners}. These configurations are more representative of industrial reactors, including gas turbines for power generation and aeropropulsion (swirl-flame stabilized methods), and afterburners (bluff body-flame stabilized methods). In both cases, the environment presents large gradients in composition and temperature, as well as is characterized by complex velocity fields. At the burner scale, different regions corresponding to the cold reactant gases, the flame, and the hot post-combustion products can be identified, with characteristic length scale that of the geometrical elements of the burner (order 1-10 cm). Note that, at the local level, the turbulent flames typically expected under these circumstances can be pictured as ensembles of laminar flame elements, or flamelets, embedded in a turbulent flow field, so that the configurations presented in figure~\ref{fig:academic_burners} might still be relevant here. 

In the plasma-assisted combustion literature both, bluff body- and swirl-stabilized burners, have been studied to demonstrate flame static and dynamic stability enhancement by plasma. For static stability studies, or extensions of the blowoff conditions by plasma-assistance, the flame environment can be mostly considered in steady state. For dynamic stability studies, or suppression of combustion instabilities by plasma-assistance, the environment is intrinsically unsteady, as it is undergoing oscillations in pressure and flame location. The electrode arrangements used primarily depend on the flame stabilization scheme: for bluff body-stabilized burners \cite{Pai2010,Pilla2006,Pham2011StabilizationDischarges} and dump combustors \cite{Kim2019Plasma-AssistedConditions}, pin-to-pin electrode configurations are typically selected; whereas for swirl-stabilized burners \cite{Barbosa2015InfluenceCombustor,Moeck2013, DiSabatino2020EnhancementDischarges,Shanbhogue2022ActivePlasmas}, a pin-to-ring/ cylinder arrangement is typically preferred, so that the plasma is left to rotate freely with the flow rather than being constrained in space.


\section{Nanosecond Pulsed Discharges in Inhomogeneous Media}\label{sec:nonuniform steady}

Whether static or dynamic, common to these environments is the presence of sharp gradients in composition, temperature, pressure, as well as flowing gas conditions. Other sources of inhomogeneity arise from the electric field; as well as the structure of the discharge itself, often developing in the form of filamentary structures. In particular, the geometry of the electrodes guides the localization of the discharge, volume of plasma activation, and possible localization in regions of high curvature, feeding into the ignition and combustion problems \cite{Mao2022_model,Bane2015}.

Under pulsed voltage, streamer breakdown is favored and several authors have studied the implications of nonuniform media in streamer development. Nonuniform environments have been considered in the context of atmospheric pressure plasma jets and streamer development along helium-air mixing layers \cite{Boeuf2013,Breden2012_3,Naidis2010,Naidis2012,Douat2011,Naidis2011,Breden2011,Guerra-Garcia2015CanEffect}; weakly ionized discharges in the upper atmosphere (such as sprites) and density gradients along the propagation length of the streamer \cite{Luque2010,Luque2012,Opaits2010,Sigmond2004,Babaeva2009,Babaeva2009_3}; streamer development in two-phase media \cite{Pillai2022,Babaeva2009_2,Bruggeman2009}; plasma aerodynamic flow control and the development of streamers with shock and rarefaction waves \cite{Starikovskiy2020}; and plasma-assisted combustion \cite{Guerra2013,Guerra-Garcia2015CanEffect,Guerra-Garcia2015a}. The impact of these nonuniformities on the breakdown conditions of NPD, as relevant to flame environments, is discussed in what follows. 

\subsection{Composition and Density Gradients}
\begin{figure}[htb]
    \centering
\includegraphics[width=0.45\textwidth, trim={0cm 0cm 0cm 0cm}, clip]{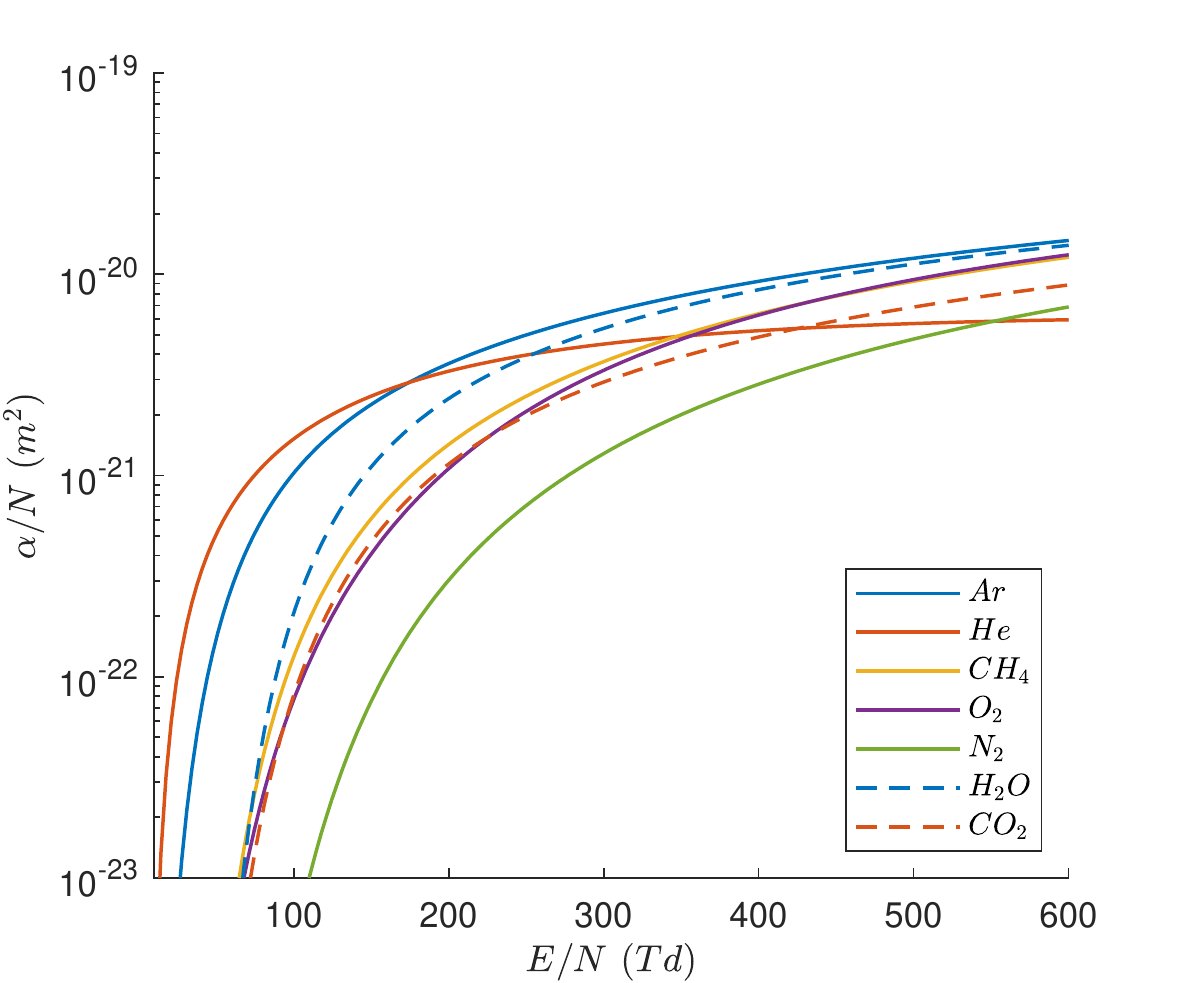}
    \caption{Townsend ionization coeffcients for major species in flame and typical dilutants used, calculated using BOLSIG+. Cross sections used are detailed in text.}    \label{fig:Townsend_graph}
\end{figure}

The first contributor to the medium inhomogeneity is the variation in the ionization properties of the gas, as affected by the local composition and gas number density. The impact of the gas composition and number density on the ionization properties is captured by the first Townsend coefficient, $\alpha$ \cite{Raizer1991,Govinda2006}:

\begin{equation}
\label{eq:Townsend-coeff}
\frac{\alpha}{N}= F \exp{\left(-\frac{G}{E/N}\right)},
\end{equation}

where $F$ and $G$ are composition-dependent. Figure~\ref{fig:Townsend_graph} shows the Townsend ionization coefficients for the major species found in methane flames, and some popular dilutants, as calculated by BOLSIG+ \cite{Hagelaar2005}. The cross sections used in the calculation are those reported in the Hayashi database \cite{lxcat_hayashi} for CH$_4$, CO$_2$ and H$_2$O, the Phelps database \cite{lxcat_phelps} for O$_2$, and the SIGLO database \cite{lxcat_siglo} for N$_2$, Ar and He. Table~\ref{table:Townsend} summarizes the $F$ and $G$ coefficients for those same species by fitting the data in figure~\ref{fig:Townsend_graph}, in the range 100-500Td, to equation~\ref{eq:Townsend-coeff}.\footnote{These values are for the pure gases. When looking at mixtures, the weighted molar average of the pure substances, known as the Wieland approximation \cite{Chantry1981}, can be misleading as the electron energy distribution function (EEDF) can deviate from that of its constituents, particularly for monoatomic/molecular mixtures \cite{Maric2005}. For consistency, the effective ionization coefficient has to be evaluated by solving the EEDF of the mixture and adding the individual coefficients of each species.}

\begin{table}[ht] 
\caption{Townsend ionization constants for major species in flame and typical dilutants used, calculated by fitting the data in figure~\ref{fig:Townsend_graph}, in the range 100-500Td, with equation~\ref{eq:Townsend-coeff}.}
\label{table:Townsend}
\centering
\begin{tabular}{ c c c c } \hline\hline
 & Gas & F [$10^{-21}m^2$] & G [$Td$] \\\hline
 Reactants & O$_2$ & 28.8 &631  \\
 & CH$_4$ & 24.3 & 550  \\
Dilutants & He & 8.0& 174\\
 & N$_2$ &24.2 &867 \\
  & Ar & 19.9  &323 \\
Products & CO$_2$ & 19.3 & 558 \\
 & H$_2$O & 28.4 & 490 \\
\hline\hline
\end{tabular}
\end{table}

 When looking at the elementary flame environments in figure~\ref{fig:academic_burners}, if a set of electrodes is placed parallel to the flame front with gap distance greater than the flame thickness, the medium can be represented by three gaseous layers in series, of different electrical properties  \cite{Guerra2013,Guerra-Garcia2015CanEffect,Guerra-Garcia2015a,Guerra-Garcia2015Non-thermalEnvironments}. In this case, the main electric field is parallel to the dominant composition and temperature gradients, and a reinforcement of the discharge in the regions where the Townsend ionization coefficient is strengthened (higher temperature or favorable composition) is to be expected. For extreme conditions, if the disparity between the electrical properties of the different layers is large enough, streamer development can be blocked by the more insulating layers \cite{Starikovskiy2020} and the discharge is confined to regions of favorable ionization. 

  \begin{figure}[htb]
         \centering
         \includegraphics[width=0.47\textwidth, trim={0cm 0cm 0cm 0cm}, clip]{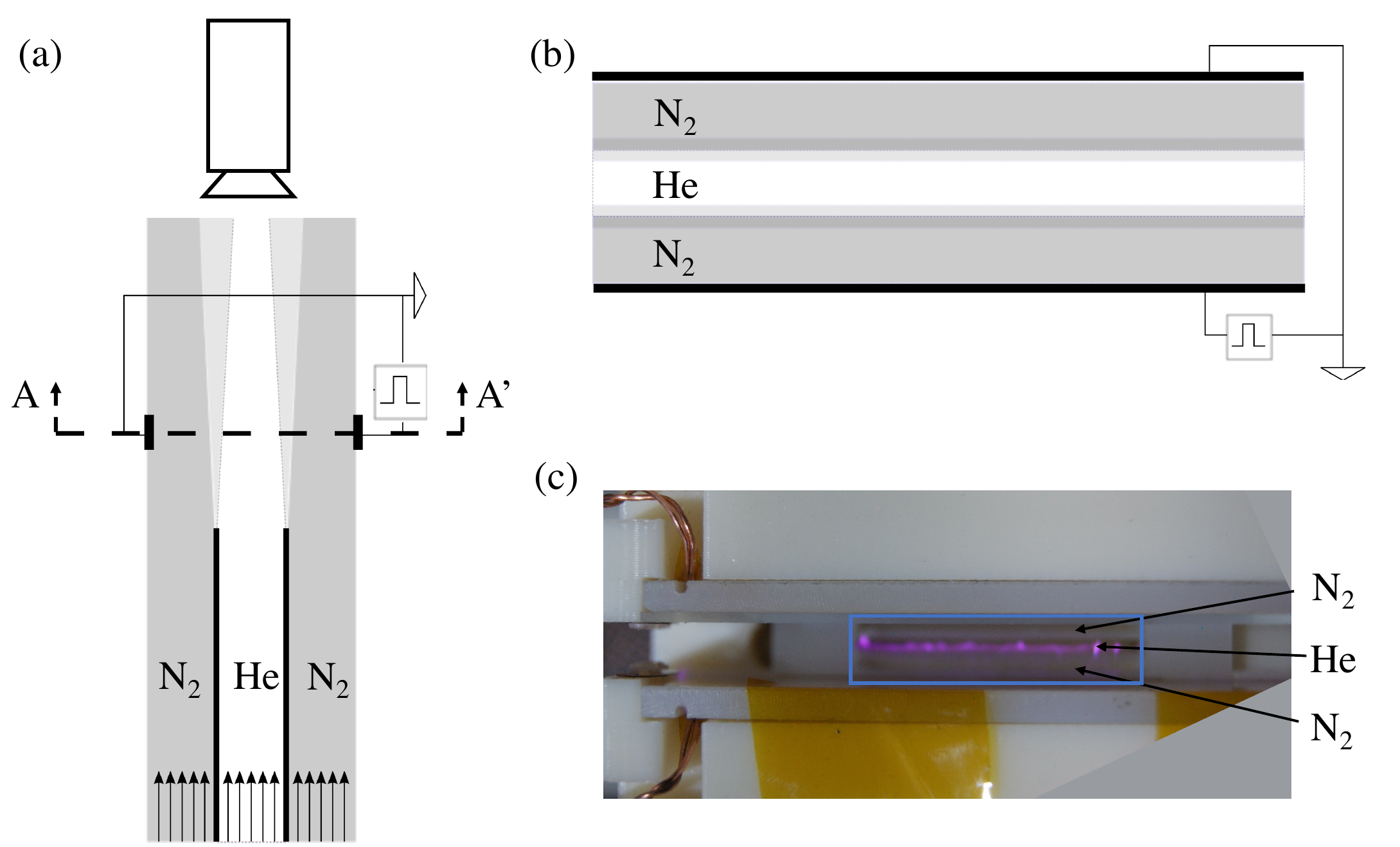}
        \caption{Gas-confined Barrier Discharge (GBD) experiments with N$_2$-He-N$_2$ parallel jets and NPD discharges. Schematics of the experiment: (a) view of three-parallel jets, (b) detail of AA' section serving as discharge cell. (c) photograph of discharge proceeding in He layer and N$_2$ layers remaining non-ionizing; blue region marks the discharge cell pictured in (b). Conditions: 6mm discharge gap (2 mm per layer), 15kV, 3kHz. The experiments are reported in Ref.~\cite{Guerra2013}.}
        \label{fig:GBD}
\end{figure}

\begin{figure}[htb]
     \centering
\includegraphics[width=0.47\textwidth,trim=0cm 0cm 0cm 0cm, clip=true]{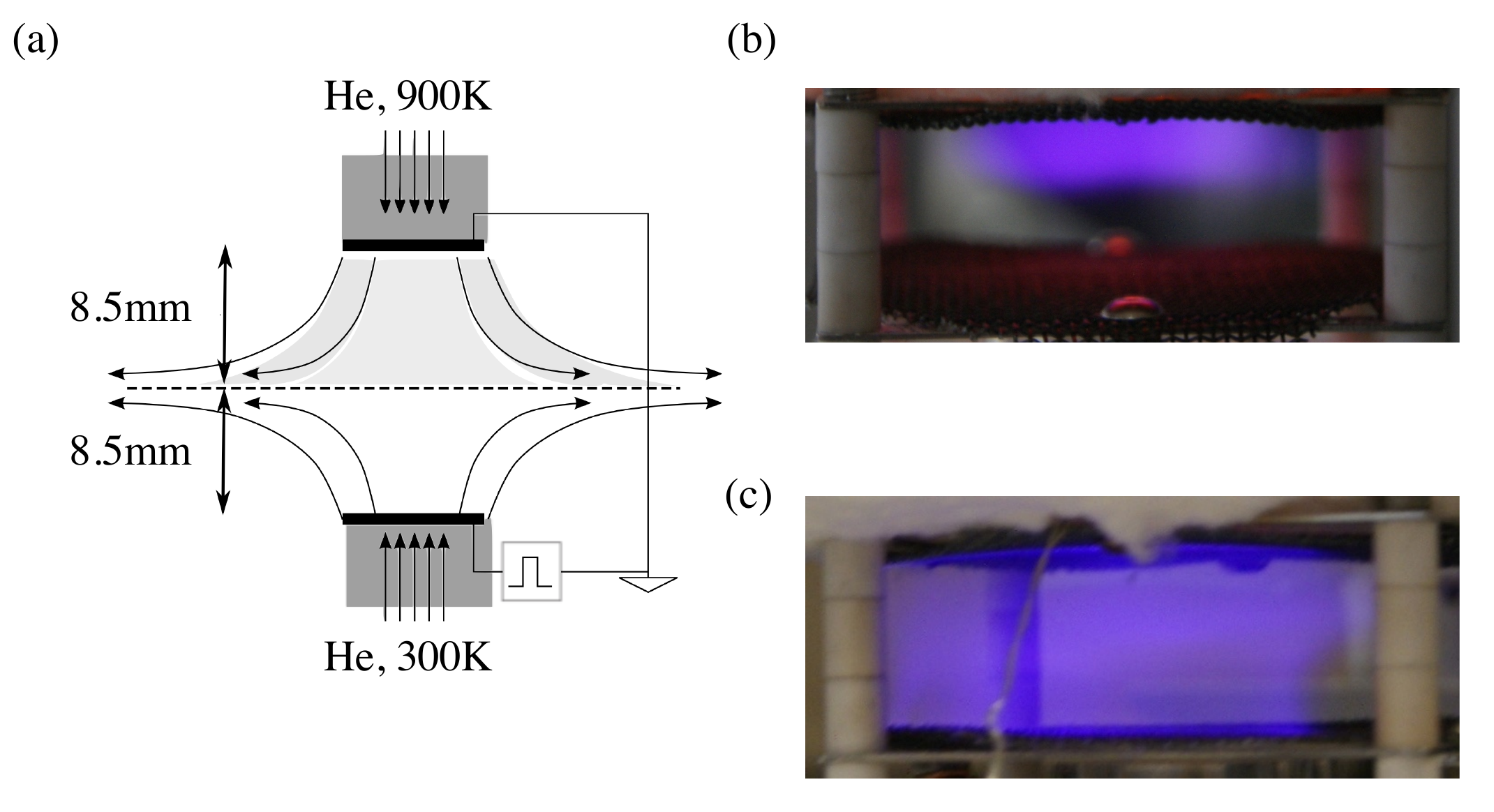}     
        \caption{Counter-flow helium jets at disparate temperatures, and atmospheric pressure. Two mesh electrodes are placed at the exit of the counter-facing nozzles. NPD discharges at 3 kHz are applied with variable voltage amplitude. (a) Schematic of the discharge cell, (b) case of selective excitation of the high temperature (low density region) for 7kV, (c) case of full breakdown of the discharge gap for 8.9kV. The experiments are reported in Ref.~\cite{Guerra-Garcia2015a,Guerra-Garcia2015Non-thermalEnvironments}.}
        \label{fig:GBD_T}
\end{figure}

 This Gas-confined Barrier Discharge (GBD) concept was demonstrated in Ref.~\cite{Guerra2013,Guerra-Garcia2015CanEffect} by applying NPD discharges across three parallel jets, with helium injected in the central stream and nitrogen injected in the side jets, figure~\ref{fig:GBD}. Helium and nitrogen were chosen for their very disparate ionization properties, as shown in figure~\ref{fig:Townsend_graph} and table~\ref{table:Townsend}. For the experimental conditions selected, the discharge proceeded in the helium stream alone and the nitrogen layers acted like a barrier to the discharge. Although the nitrogen layers are non-ionizing, they still allow for transport of charged species, including convection, electric drift, and diffusion. Alternatively, selective ionization when using NPD discharges can be realized by having a region of reduced local density, N. This situation was explored in Ref.~\cite{Guerra-Garcia2015a,Guerra-Garcia2015Non-thermalEnvironments} by applying NPD pulses across two opposed jets of different temperature, figure~\ref{fig:GBD_T}. In the experiment, the applied voltage amplitude was gradually increased to observe a transition between: a discharge confined to the hot jet, figure~\ref{fig:GBD_T} (b), and breakdown of the full gap, figure~\ref{fig:GBD_T} (c). 

Whether or not breakdown can be sustained is only one piece of the puzzle. Much of the work in the field has centered on the role of composition on the regime transitions observed, including uniform to filamentary modes. In particular, certain combustion products can promote filamentarization of the discharge (e.g., NO), and this has been attributed to their different ionization properties \cite{Adamovich2009, Gadkari2017, Rousso2020}; alternatively, certain species can have a stabilizing effect (e.g., water vapor), attributed to higher attachment rates \cite{Zhong2019}. The role of heat release or absorption by chemical reactions in the transition between uniform to filamentary regimes, the so-called thermo-chemical instability mechanism, has also been a matter of study \cite{Zhong2019, Rousso2020}. 

\subsection{Flow Fields}\label{sec:flow_fields}

Aside from the local state of the gas (pressure, temperature, composition), the flow field within the discharge volume can play a major role in the discharge dynamics, controlling aspects such as the discharge regime encountered or the spatial structure of the energy deposition. 

Several studies by Leonov et al.~\cite{Leonov2016,Leonov2008,Leonov2014,Houpt2017} have considered the influence of airflow on so-called near-surface electric discharges, including dielectric barrier discharges (DBDs), pulsed DC discharges, and quasi-DC discharges, mostly in the context of aerodynamic flow control and supersonic conditions. The two-way interaction between flow and discharge is well exemplified by the case of quasi-DC discharges. Despite the DC voltage, the discharge is pulsating and the transient nature of the filamentary discharges, with temperatures of about 3000K moving with the flow downstream until they split and reattach to a new location \cite{Leonov2008}, is the result of a strong coupling of the moving gas and the discharge dynamics \cite{Houpt2017}. 

\begin{figure}[htb]
    \centering
    \includegraphics[width=0.45\textwidth]{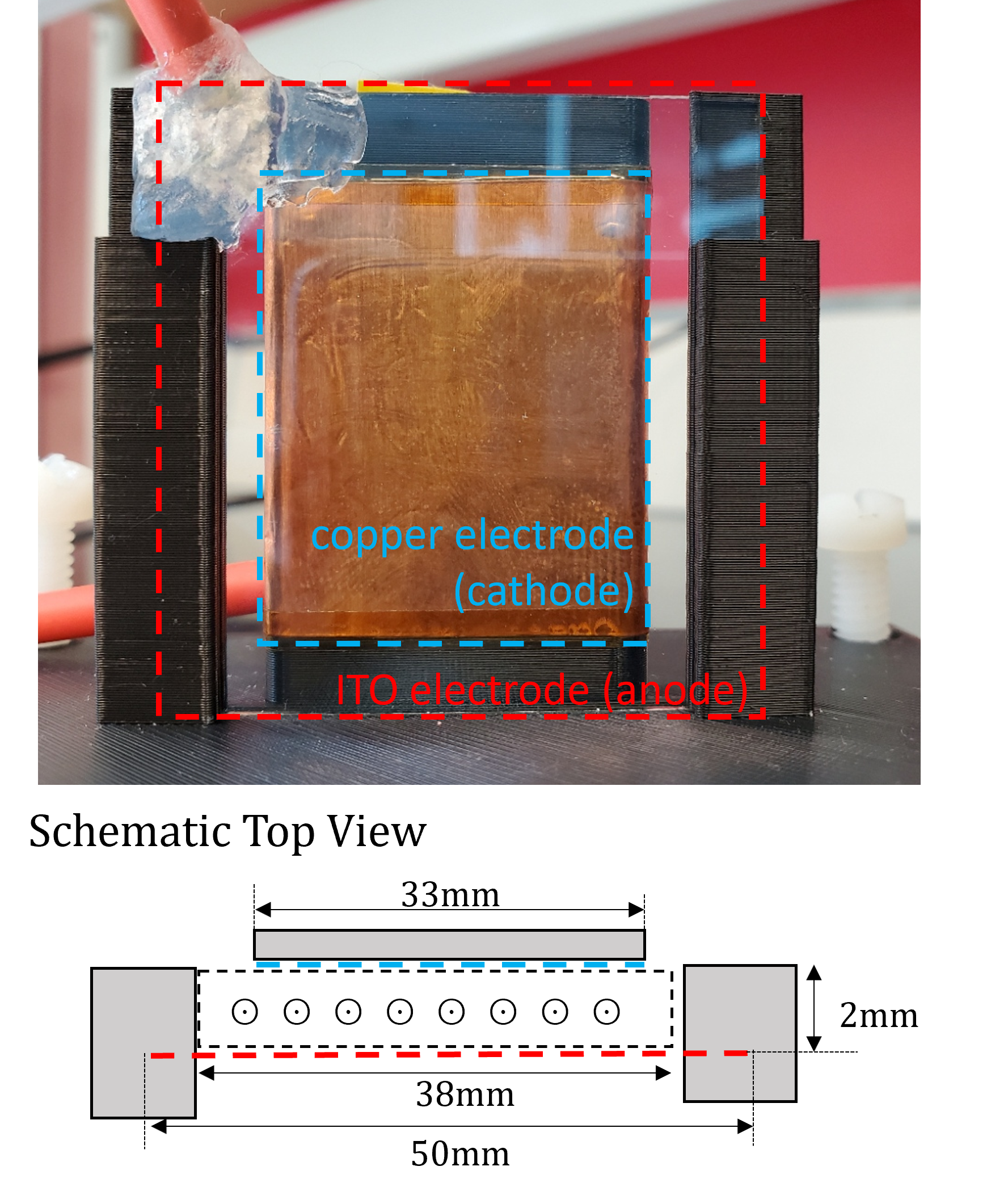}
    \caption{(Top) Image of the experimental setup seen from the perspective of the camera. The ITO electrode is transparent; its border is indicated by the dashed red line. (Bottom) Top view showing the flow inlet location and open sides of the test section. Flow is out of page as indicated by the $\odot$ symbol.}
    \label{fig:ITO_setup}
\end{figure}

\begin{figure*}[htb]
    \centering
\includegraphics[width=1\textwidth]{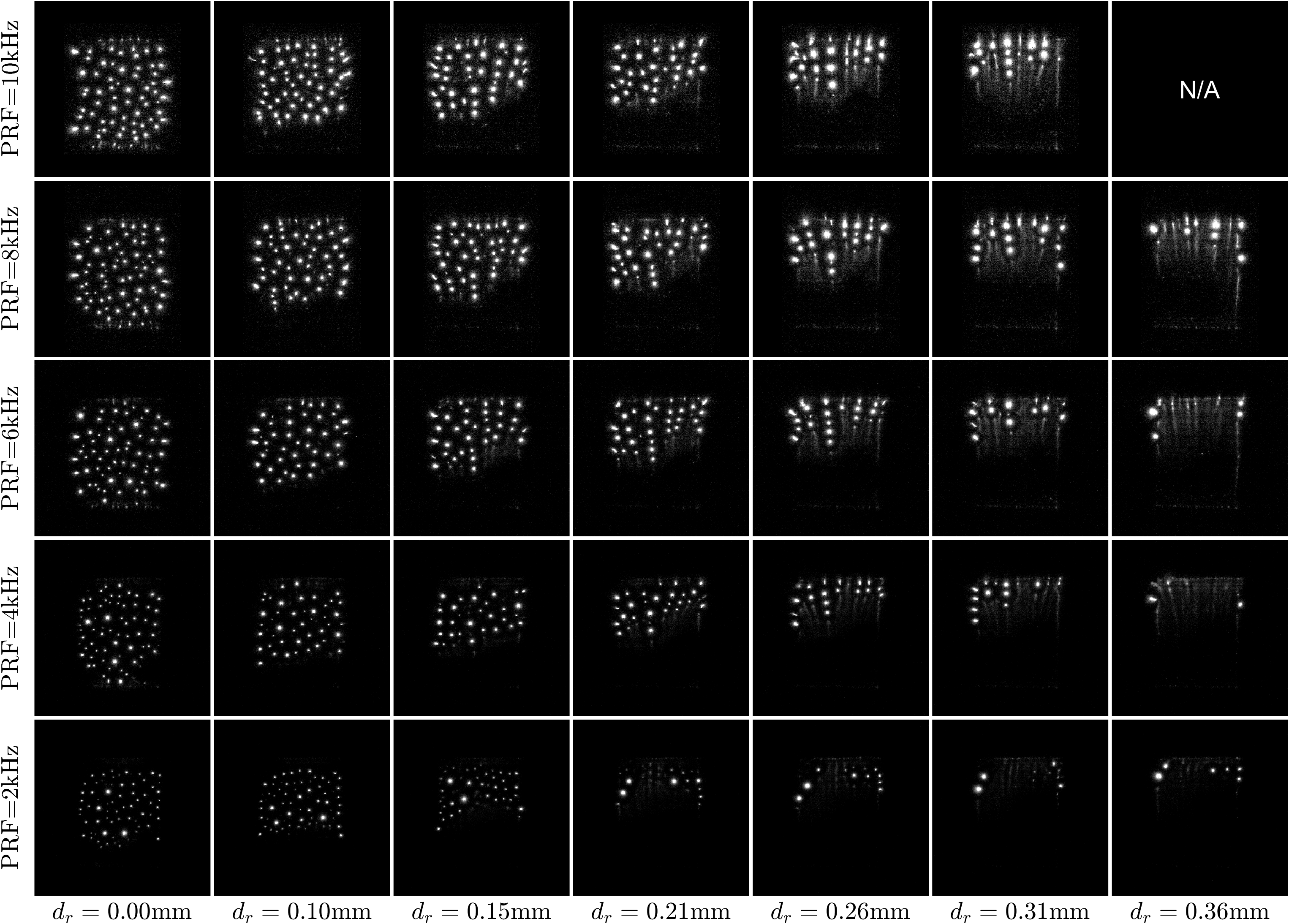}
    \caption{Head-on visualization of the discharge as a function of pulse repetition frequency, PRF, and a characteristic flow replenishment distance, $d_r$ (ratio of the inlet speed-to-PRF). The different replenishment distances are obtained by varying the mass flow rate in the reactor at fixed PRF. Each image corresponds to the time-integrated intensity of pulse 501 in a train with fixed applied voltage of 18 kV. Flow going up. The experiments are reported in Ref.~\cite{Guerra2023_scitech}. The view direction shown in these images is parallel to the applied electric field and the discharge is observed through the clear ITO anode. Each image shows the time-integrated visible-spectrum light from a single pulse. Camera settings were kept fixed for each PRF, but camera sensitivity and frame rate were varied as PRF was changed; relative intensity can be compared between columns but not between rows.}
    \label{fig:flow_ITO}
\end{figure*}

When using NPD, bulk flow will affect the synergistic effect of consecutive pulses, relating the initial conditions of a pulse (electron densities, temperature, excited species densities, etc) to the preceding one. Filamentary discharges have also been observed to be advected by the flow in this case. The experiment reported in Ref.~\cite{Guerra2023_scitech} using a dielectric barrier discharge, DBD, reactor in flowing air at atmospheric pressure and ambient temperature illustrates these points. The reactor used a transparent anode, consisting of a 1 mm thick glass slide with a thin film of indium tin oxide (ITO) deposited on the dry-side. The cathode was made of regular copper tape. The choice of electrodes allows to take head-on images of the discharge. The ITO anode is placed orthogonal to the camera's line of sight, so that the view of the discharge is through the ITO anode and co-linear to the dominant electric fields and the propagation direction of the filamentary discharges that may form. This setup is shown in figure~\ref{fig:ITO_setup}. Another peculiarity of this rectangular reactor was that gas was injected from one side and the three other sides were left open to the atmosphere, resulting in a 2D non-uniform flow field with diverging streamlines. The setup was used to explore the effects of flow residence time, compared to the inter-pulse period, by varying the mass flow rate (or inlet speed) and the pulse repetition frequency. The key finding of the study is shown in figure~\ref{fig:flow_ITO}. A characteristic flow replenishment distance, $d_r$, defined as the ratio of the inlet speed to the pulse repetition frequency, was identified as the parameter controlling the discharge dynamics, rather than the inlet speed or the pulse repetition frequency alone. This distance represents the characteristic distance travelled by a fluid parcel between two consecutive pulses. For $d_r$ below some critical value order $\sim 100 \mu m$, microdischarges are favored; for larger $d_r$ the discharge is no longer visibly observed or becomes uniform. The transition is gradual and can be spatially resolved within the reactor due to the diverging flow field structure. As $d_r$ increases, a decreasing region of micro-discharges, downstream of the transition zone where the speed is highest, is observed. The boundary of the transition zone moves downstream with $d_r$ (either increasing inlet speed or decreasing pulse repetition frequency). Interestingly, the critical replenishment distance for the transition seems to be comparable to typical streamer radii at these conditions \cite{Pavan2020InvestigationsModels}, suggesting that filamentarization is more likely when there is a stronger synergy between pulses and any instability present will be more likely to get amplified.

\subsection{Flame Front and Reaction Zone Environments}

In the previous examples, the different contributors to the medium non-uniformity were to some extent isolated to evaluate their separate impact on the breakdown phenomena. Even the simplest flame environments depicted in figure~\ref{fig:academic_burners} present all these effects simultaneously, in addition to a pre-ionization region contained within the high-temperature zone \cite{Lawton1963}. For hydrocarbon flames, ions and electrons are generated in the reaction zone of the flame in amounts much in excess of thermodynamic equilibrium values as a result of chemi-ionization reactions, in which species undergo a chemical rearrangement that is sufficiently exothermic to ionize one of the products \cite{Lawton1967,Semenov1970,Pedersen1993,Starik2002}. The most important reaction contributing to the formation of ions and electrons is: 
\begin{equation}\label{eq:chemi}
\text{CH} + \text{O} \longrightarrow \text{CHO}^+ + e^-,
\end{equation}
where the CH is either in ground state or electronically excited \cite{Cool1984}. Although CHO$^+$ is the precursor ion, the dominant ion is usually H$_3$O$^+$. The region of the flame containing charged species is comparable in size to the reaction zone thickness, e.g., for stoichiometric premixed methane-air flames this thickness is $\delta_i\sim$0.6mm \cite{McLatchy1979,Ju2004}. Note that this thickness may vary with the flame parameters, but will be of the order of a fraction of a mm, and much less than the thickness of the high temperature region. The ionization fraction within this region varies in the range $n_i/N\approx10^{-8}$-$10^{-7}$, for methane-air flames, peaking close to stoichiometric conditions (slightly fuel rich)   \cite{Calcote1962}. The presence of these chemi-ionization electrons and ions impacts the way the electrical energy is coupled to the gas in a number of ways.

First, this region of pre-ionization is critical when using sub-breakdown fields. For DC and AC waveforms, the chemi-ionization ions are responsible for the generation of ion-driven winds, a phenomenon that has been studied since the 1960s \cite{Lawton1968,Lawton1967,Guerra-Garcia2015}. When using sub-critical microwave (MW) energy, the energy can be directly coupled to the chemi-ionization electrons, which are heated up selectively \cite{Ju2004,Sullivan2004,Stockman2009_phd,Michael2012}. 

\begin{figure}[htb]
    \centering
\includegraphics[width=0.45\textwidth, trim={0cm 0cm 0cm 0cm}, clip]{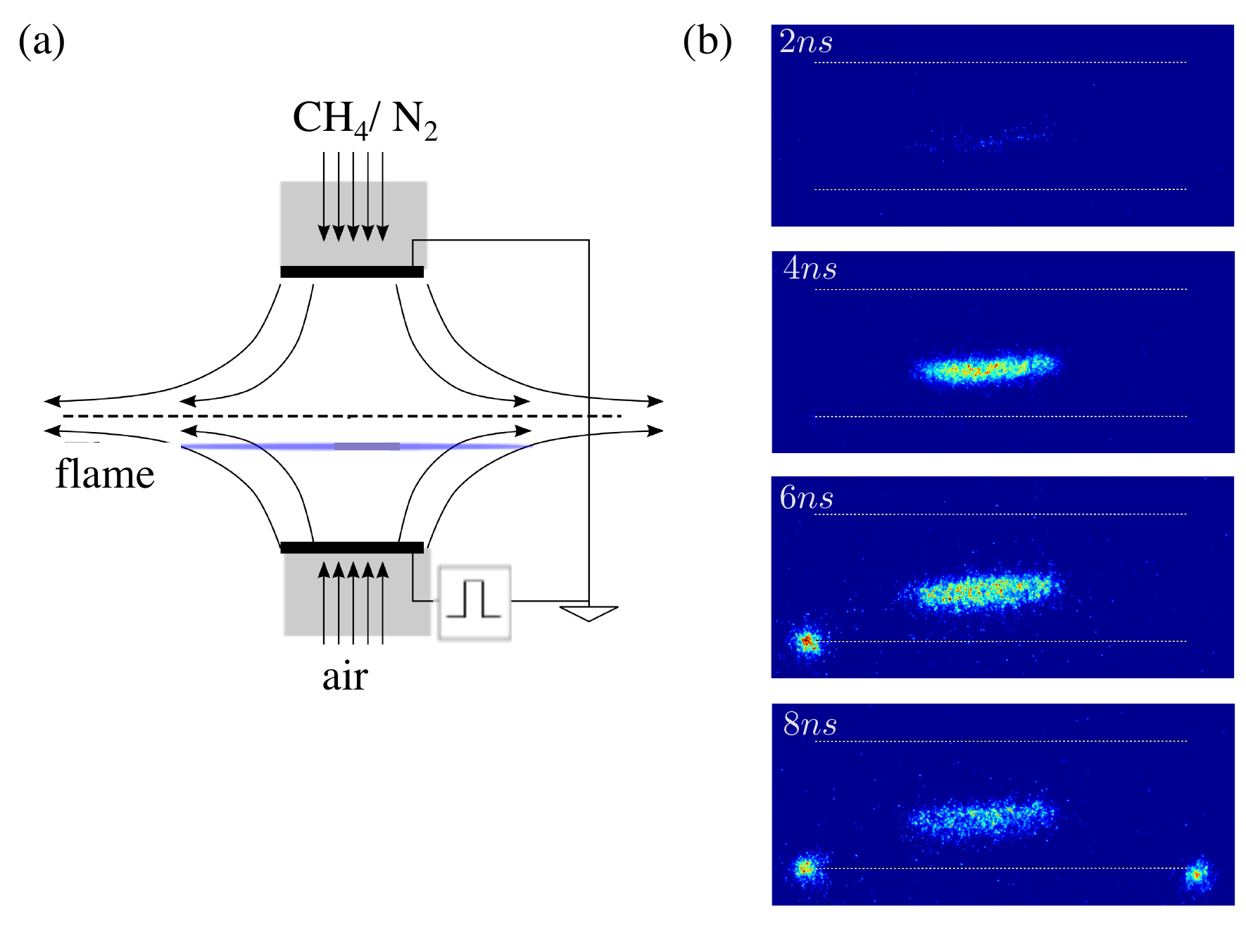}
    \caption{Plasma emission coupled to flame front in a counterflow nonpremixed burner. (a) Discharge cell schematic. (b) Timeline of discharge evolution during voltage pulse application; time stamps correspond to time with respect to NPD pulse onset, and white-dashed lines indicate location of the electrode meshes located at the exit of the counterflow nozzles. Images correspond to pulse number 1000 in a train and the exposure time is 0.5ns. The plasma luminosity is contained within the flame and the edges of the powered electrode. The experiments are reported in Ref.~\cite{Guerra-Garcia2015a}.}
    \label{fig:nonpremixedNPD}
\end{figure}

In the case of NPD, the presence of the chemi-ionization electrons has several implications. They contribute to the sustainability and repeatability of the discharge: typically the NPD strategy relies on the use of kHz frequencies, chosen to match the recombination times of electrons, in order to sustain a minimum electron population at all times \cite{Pai2009}. In the case of a flame front, a baseline level for the electrons already exists, with magnitude comparable to the minimum electron mole fractions measured between pulses in glow-like NPD discharges~\cite{Kruger2002}. Pre-ionization strategies can also be used to improve the uniformity of the discharge \cite{Palmer1974,Levatter1980}: electron avalanches can overlap, smoothing out local gradients, and favoring a homogeneous breakdown condition. Huang et al.~\cite{Huang2014} discussed the role of pre-ionization in the breakdown properties of the gas in the context of NPD discharges at different pulse repetition frequencies. Their experiments of NPD discharges, at atmospheric pressure using pin-to-pin electrodes in Ne and Ar mixtures and variable frequency in the 5-667kHz range, showed a reduction in the breakdown voltage for the highest frequencies tested, which they attributed to the high residual charge density between pulses. Subsequent experiments with Fast Ionization Waves by the same team~\cite{Huang2018}, proposed that higher pulse repetition frequencies also lowered the degree of non-equilibrium of the discharge. Note that a lower local breakdown threshold at the flame front location would lead to the coupling of the discharge to this region. Such a selective excitation of the flame to the flame front region, by NPD plasma, is presented in Ref.~\cite{Guerra-Garcia2015a} using a counterflow nonpremixed methane/air burner with mesh electrodes parallel to the flame, figure~\ref{fig:nonpremixedNPD}. The photographs shown in figure~\ref{fig:nonpremixedNPD}(b) were taken by a PicoStar HR12 (LaVision GmbX) intensified charge-coupled device (ICCD) camera using a gate of 0.5ns and correspond to the integrated emission over a broad wavelength range. Within an applied voltage range, the luminosity of the plasma was fully contained within the flame front region, with no contact with the electrodes, suggesting localized energy deposition to the flame. 

\section{Nanosecond Pulsed Discharges in Dynamic Environments}
\label{sec:dynamic}

Much of the work in the field exploits the repeatability of the pulses and centers on the characterization of the discharge once a quasi-periodic behavior has been reached \cite{Minesi2022}. This occurs after a few tens or hundreds of pulses, if the medium is not significantly changing at the timescale of the applied train of pulses. For operation at a few kilohertz, e.g. 10 kHz, and trains of $\sim$100 pulses, this timescale is of the order of $\sim$10 ms. Many combustion processes of interest cannot be considered steady within that time-frame. E.g., plasma-assisted ignition is greatly influenced by inter-pulse coupling considerations; a laminar methane-air premixed flame will travel about 4 mm in 10ms, which is typically larger than the flame thickness; and a combustor presenting an instability with a frequency of oscillation of 100 Hz will have undergone a full cycle of variation when pulse 100 is applied. For these cases, and also when considering turbulent flows, a steady-state pulsing behavior will not be encountered, and individual pulses will need to be studied separately to have the complete picture of the interaction. Three examples are reviewed in what follows.   

\subsection{Inter-Pulse Coupling and Cumulative Effects of Pulses}

In otherwise uniform conditions, the application of a train of NPD discharges introduces a modification to the medium as each consecutive pulse is applied. In this situation, individual pulses are triggered in different background conditions (in terms of temperature, composition, and pre-ionization), as modified by the prior pulse. One of the first works reporting the cumulative effects of pulses was Ref.~\cite{Pancheshnyi2006}, who measured the properties of individual pulses in a train, in terms of energy deposited, temperature increase, and optical spectra; both in atmospheric pressure air and in a plasma-assisted ignition experiment with variable pressure and propane/air mixtures. The study revealed that during a train of 10 pulses in air, the first few pulses had negligible conduction current and energy limited to around a hundred microjoules, whereas the last pulses in the train delivered about ten times more energy (order mJ). The difference was attributed to the pool of electrons and other active species (radicals, excited molecules) building up, as well as the cumulative heating effects of the pulses, leading to more favourable conditions for the discharge for the later pulses in the train. The number of pulses for the transition depends on the state of the gas, as well as the electrical conditions selected, and plays an important role in plasma-assisted ignition studies.

The implications of the cumulative effects of pulses on the ignition probability of a flowing mixture was studied by Lefkowitz and Ombrello in Ref.~\cite{Lefkowitz2017}. The experiments varied a number of parameters including total energy deposition, inter-pulse time, equivalence ratio, gap distance, and flow velocity; and identified three regimes of inter-pulse coupling: fully-coupled, partially-coupled, and decoupled. The fully-coupled regime occurs at the highest pulse repetition frequencies (for given flow residence times), and in this regime all the pulses are deposited in the same small volume, resulting in high energy density, and a high probability of developing an ignition kernel that can freely propagate into a flame. For the decoupled regime, occurring at the lowest pulse repetition frequencies (for given flow residence time), the ignition kernels do not interact until they have expanded significantly, so that the total ignition probability is proportional to the ignition probability by a single pulse. Refs.~\cite{Lefkowitz2017,Mao2022_interpulse} exemplify the complexity of optimizing a plasma-assisted ignition and combustion problem, given the interactions between plasma, flow, and combustion.     

\subsection{Transient Laminar Flame Passage}\label{section:transientflame}

A simple platform to visualize the impact of a transient combustion environment on the discharge behavior is provided in Ref.~\cite{Pavan2022}. In this work, the evolution of a DBD discharge under NPD voltage during passage of a laminar premixed methane-air flame, in a narrow channel, is presented. For a discharge developing in this background gas, all of the effects considered in section~\ref{sec:nonuniform steady} will be present, and in addition every pulse applied will see a slightly different environment. 

\begin{figure*}[htb]
     \begin{subfigure}[t]{1.0\textwidth}
        \centering
        \includegraphics[width=1\textwidth, trim={0cm 8cm 0cm 7cm}, clip]{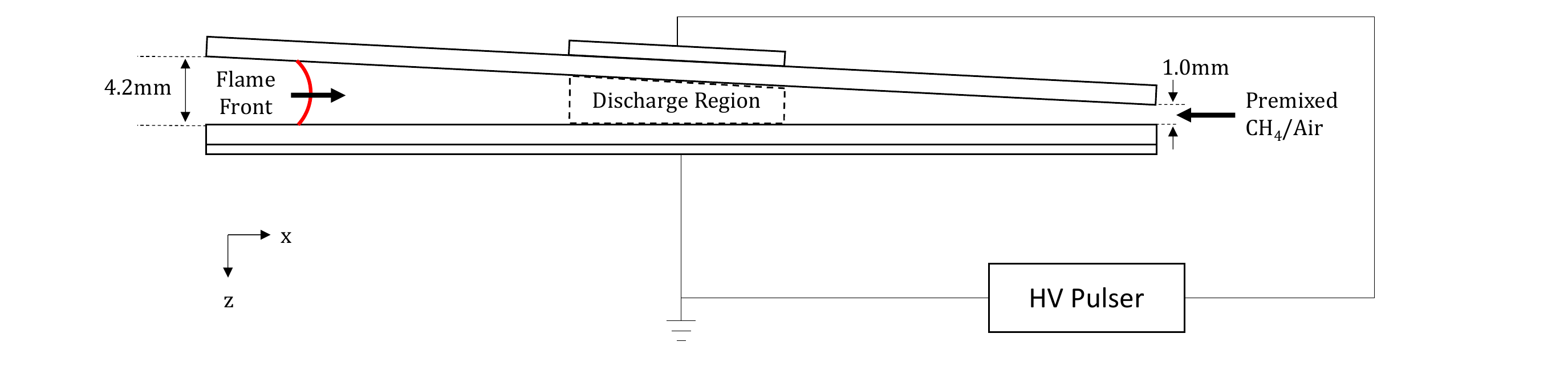}
        \caption{Setup for mesoscale flame experiments in tapered channel. Cathode is 20cm long, extending the entire length of the test section; anode position and length are variable. Test section is 3cm wide in the y-direction. }
        \label{fig:tapered_test_section}
    \end{subfigure}
    \begin{subfigure}[b]{0.55\textwidth}
        \centering
        \includegraphics[width=1\textwidth, trim={0cm 0.5cm 0cm 0cm}, clip]{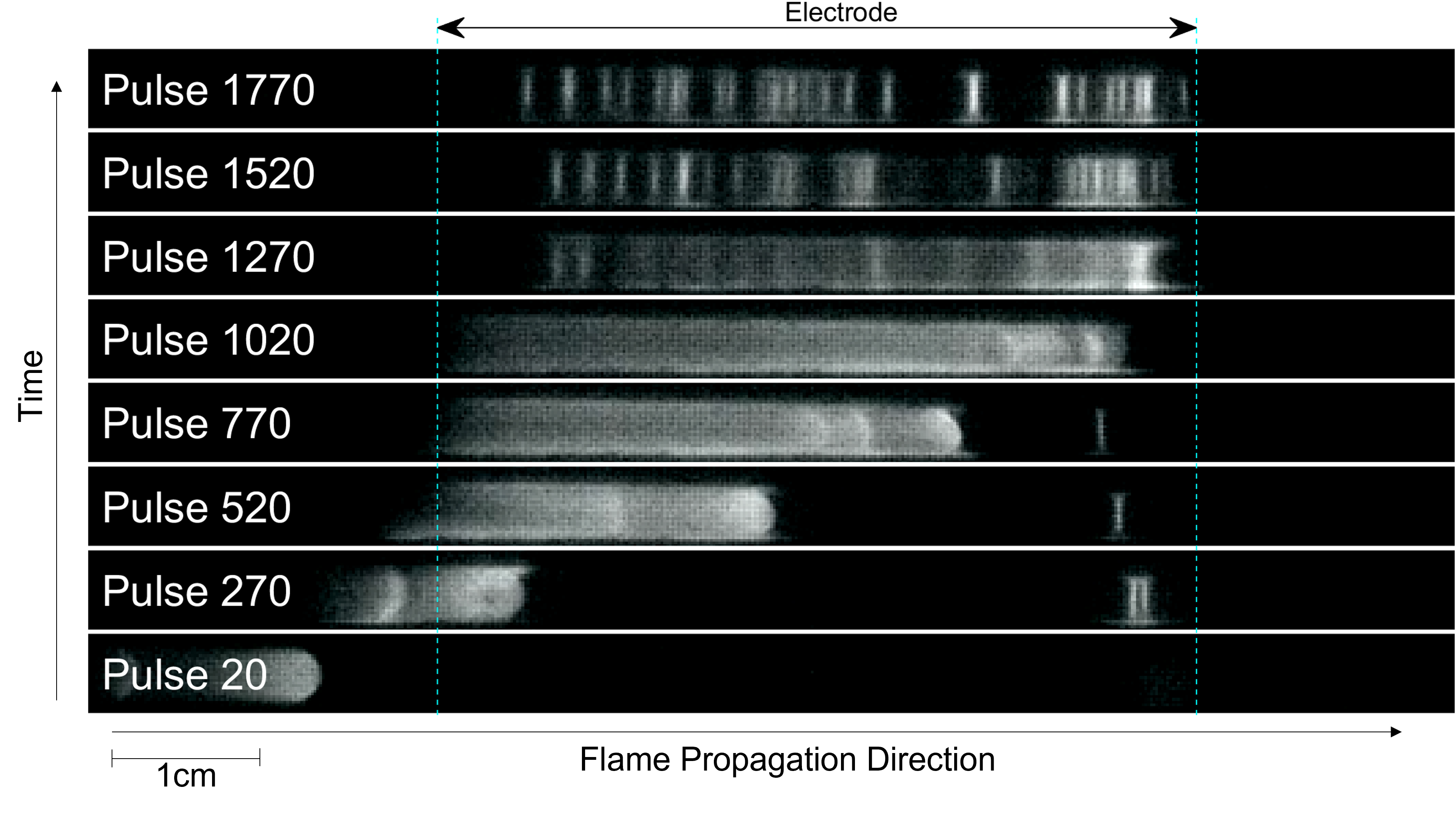}
        \caption{Direct imaging of individual pulses/ flame interaction.}
        \label{fig:regimes_flame_passage}
    \end{subfigure}
    \begin{subfigure}[b]{0.35\textwidth}
        \centering
        \includegraphics[width=1\textwidth, trim={0cm 0cm 0.5cm 0cm}, clip]{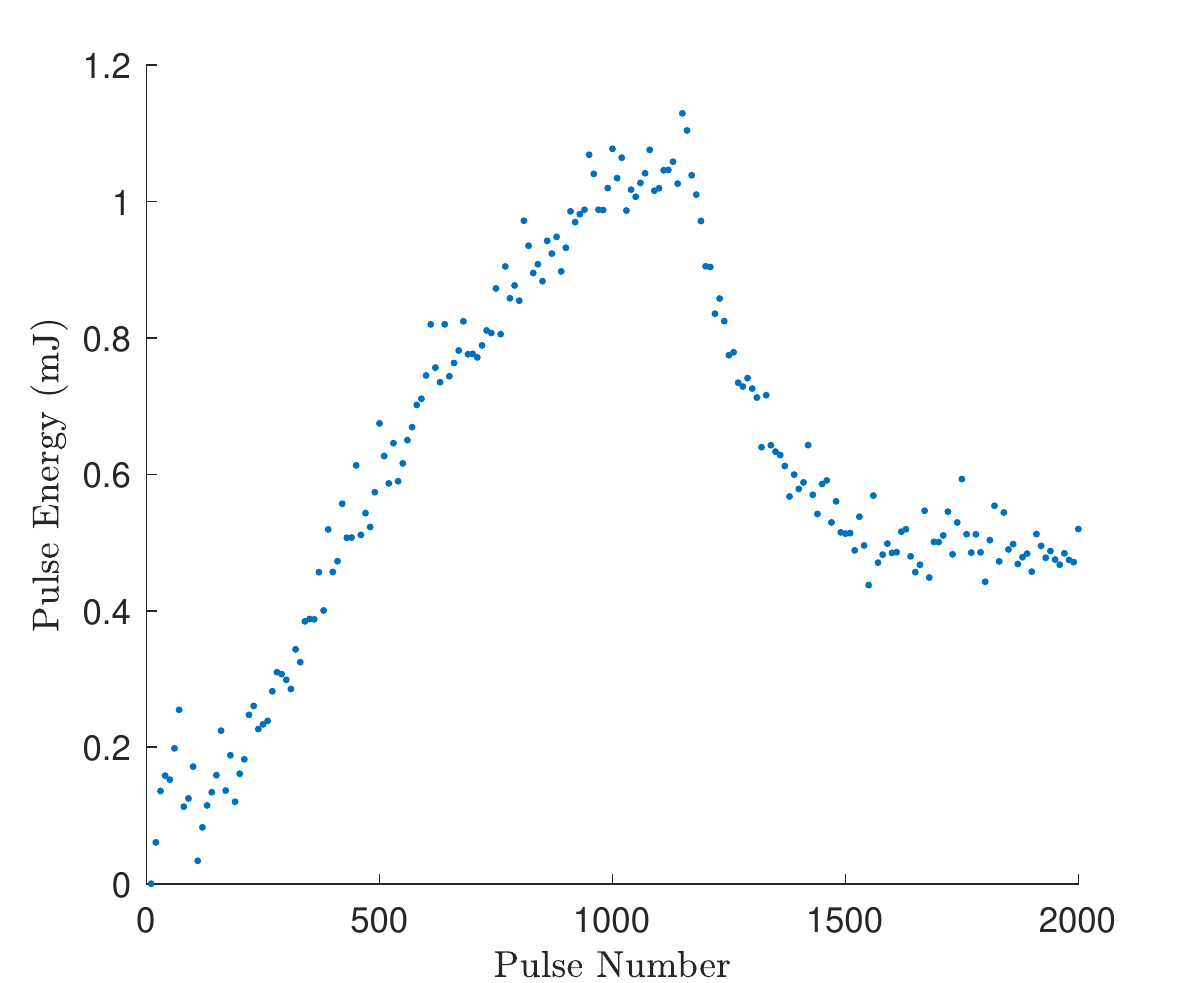}
        \caption{Per-pulse energy measurements.}
        \label{fig:nrg_evo}
    \end{subfigure}
    \caption{(a) Experimental setup and (b,c) sample results for experiments in tapered channel with laminar flame. Conditions for results shown: f=8kHz, V=14.7kV, $\mathrm{\phi}$=1.1, Q=200sccm, camera frame rate 2kfps. The discharge transitions from a uniform regime to micro-discharges, with consequential modifications to the spatial structure of the energy deposition, as well as total energy delivery. The experiments are reported in Ref.~\cite{Pavan2022}.} \label{fig:laminarflame}
\end{figure*}

Figure~\ref{fig:laminarflame} shows a schematic of the experimental setup and presents an example of the discharge evolution, in response to flame passage, for a case of pulses at f=8kHz, V=14.7kV and flow conditions of $\mathrm{\phi}$=1.1, 200sccm. The images shown time-integrate the visible spectrum light from 4 sequential pulses and were taken using an Edgertonic SC2+ monochrome high speed camera. To distinguish flame from discharge, the experiment was also performed with a camera frame rate twice the discharge frequency; an example of this is shown in the original work~\cite{Pavan2022}. The first few pulses (pulse 20) are applied before the flame enters the inter-electrode gap, and no visible discharge is appreciated. As the flame enters the inter-electrode gap (pulse 270), a uniform plasma is ignited in the hot burnt gas behind the flame, and that uniform plasma region increases in size as the flame continues to propagate (pulses 520, 770, 1020). During this time of plasma size increase, the energy deposition increases close to linearly with pulse number (or discharge size extension), figure~\ref{fig:nrg_evo}. An inflection point is seen around pulse 750, when the leftmost edge of the discharge detaches from the electrode (the cooling of the gas at the downstream side of the flame cannot support breakdown in that region anymore). Between pulse 1020 and pulse 1270, the flame reaches its furthest most position and is quenched before it exits the inter-electrode region, as a result of adverse interactions with the plasma and the tapered walls of the test cell \cite{Pavan2022}. This point also coincides with the peak energy deposition. At this point, the heat source keeping the gas hot is extinguished and the discharge is seen to transition into a microdischarge regime (pulse 1270, 1520), which persists until the pulses are turned off (pulse 1770). After flame extinction, the per-pulse energy rapidly decays and remains approximately constant while the micro-discharges persist. The test section was constructed with a shallow taper to study plasma interaction with flames near the quenching distance, and the extinction of the flame is partially caused by the narrowing walls of the channel. This decreasing gap size also causes the filaments to form at the right side of the electrode (pulses 270, 520). Additionally, the plasma has been observed to cause premature quenching of the flame, as reported in Ref.~\cite{Pavan2021PlasmaFlames}. While the precise mechanism is not currently known, a possible cause is discharge-produced flow disturbances predicted by subsequent modelling work~\cite{Pavan2023}.

This example illustrates that the combustion environment is driving the volume of gas activated by plasma, the energy delivery, as well as the plasma regime encountered. The effects discussed for static environments are present, e.g., selective ionization of hot gas regions and regime transitions, with the added complexity of the pulse-to-pulse variability of the discharge. This has important implications when considering plasma actuation strategies. Any control scheme will likely be based around a particular discharge mode and expected energy addition to the gas, however, a changing gas environment driven by combustion dynamics can lead to different behaviour. The bench-top scale platform here is useful for probing the mechanisms of interaction in a way not possible in complex environments of more practical relevance, one of which will be discussed in the following section. The mesoscale burner will be revisited in section~\ref{sec:new_results}. 

\subsection{Swirl-Stabilized Burner Undergoing Combustion Dynamics}

\begin{figure}[htb]
    \centering
\includegraphics[width=0.45\textwidth, trim={0cm 0.5cm 0cm 0.5cm}, clip]{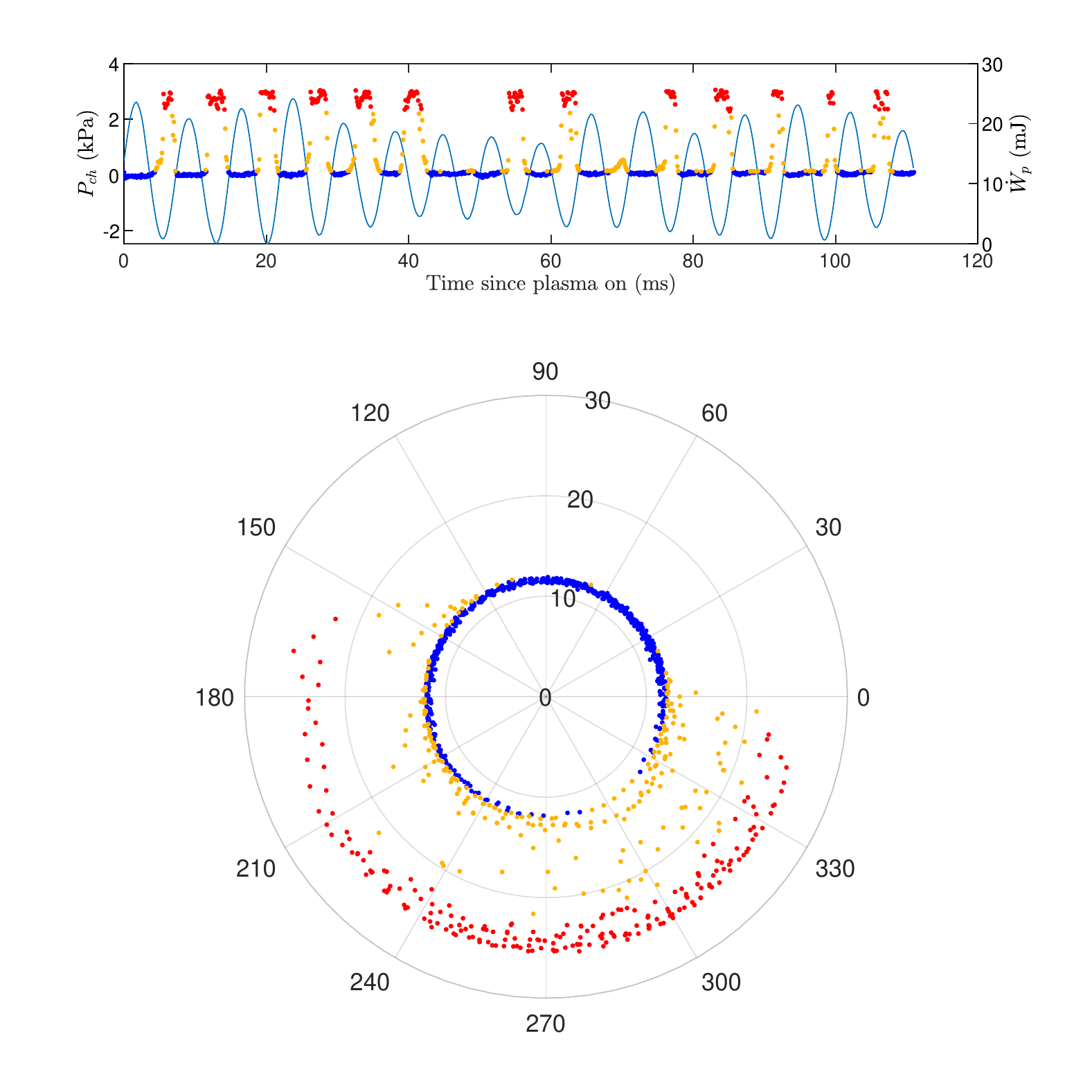}
    \caption{Correspondence between energy deposition by NPD discharge and pressure oscillation cycle during thermoacoustic instability. (Top) Sample time-trace. (Bottom) Polar plot of the discharge energy, radial coordinate in mJ, as a function of pressure phase angle. Blue markers are streamer corona, red markers are nanosecond spark, orange markers are transitional. Conditions $\phi=0.78$, PRF=$9kHz$, $V_{peak}$=12.4kV. Experiments are reported in Ref.~\cite{Pavan2022_aviation}.}
    \label{fig:swirl_combustor}
\end{figure}

Whereas the experiment discussed in the previous section represents an academic platform, the high level observations should also translate to the industrially-relevant environments discussed in section~\ref{sec:industrial}. In particular, such a strong coupling of the discharge behavior to transient flame environments could have important implications when using NPD plasmas to stabilize lean flames and suppress thermoacoustic instabilities, section~\ref{sec:intro}. Such an environment has been studied in Refs.~\cite{Shanbhogue2022,Pavan2022_aviation} and is briefly summarized in what follows. The studies consider a premixed swirl-stabilized burner described in detail in Ref.~\cite{Shanbhogue2022}, which presents combustion dynamics with a dominant frequency around 120 Hz at the lean equivalence ratios tested. The reactor operates at atmospheric temperature and pressure with lean methane/air mixtures, and the operating flame power is around 14 kW. The flow is in this case turbulent. The electrode system is provided by FGC Plasma Solutions and consists on a center-body pin-anode inserted into the swirler hub, to which NPD discharges are applied; the cathode is the annular body of the injector section. The NPD plasma was shown to provide significant attenuation of the pressure oscillations \cite{Shanbhogue2022}, and the discharge was seen to be strongly affected by the combustion dynamics \cite{Pavan2022_aviation}, as illustrated in figure~\ref{fig:swirl_combustor}. In this plot, the pressure trace is used to track the combustion dynamics, and should not be interpreted as the driver of the plasma-regime modifications (the change in pressure is at most $\sim 4\%$ of the ambient value). Rather, it is correlated to the heat release rate and the position of the flame with respect to the central electrode, which are the more likely drivers of the discharge evolution. Under these conditions, during the instability cycle, three different regimes were identified, corresponding to different energy depositions and visual appearances: (i) a high-energy ($\sim 20 mJ$) ns-spark; (ii) a low-energy ($\sim 10 mJ$) streamer corona\footnote{In the streamer corona mode, most of the energy is not deposited in the corona itself but rather upstream of the electrode via parasitic resistance and discharge, see \cite{Pavan2022_aviation}.}; and (iii) a transitional regime.

The high-energy regime appears within a specific region of the pressure phase cycle, coinciding with conditions where the flame luminosity is at its maximum (flame strengthened). The lower-regime appears at the opposite region of the cycle, when the flame is weakest. The transition between both modes is gradual, with pulses depositing intermediate energy levels in between. In this situation, the dynamic combustion environment is controlling the discharge energy deposition process, with possible implications on its ability to suppress the instability, if not adequately accounted for.


\section{Coupling of the Forward and Backward Problems: Imaging Study of a Mesoscale Flame with Plasma}
\label{sec:new_results}
To further emphasize the two-way discharge-flame interaction, the experiments reviewed in section~\ref{section:transientflame} are extended in this perspectives article to include a novel visualization enabled by the use of transparent ITO electrodes. By applying the NPD to an ITO anode, the interaction between flame and plasma can be visualized in 2-dimensions, revealing features that were previously hidden. The experimental setup used is the same as that described in Ref.~\cite{Pavan2022} and shown in figure~\ref{fig:tapered_test_section}, and is here summarized for completeness, including the new elements added. The reactor is a tapered quartz channel with a width of 3cm and a height varying from 4.2mm to 1mm over 20cm (the height changes by less than 0.2mm over 1cm). Premixed methane-air, at an equivalence ratio of $\phi=1.06$, enters from the small end of the taper and flows towards the large end, the flow rate is set to 400sccm. At the exit of the reactor, a spark plug is used to ignite the mixture. The flame then flashes back through the channel and is thermally quenched by the narrowing walls of the chamber. The 2mm quartz walls of the reactor are used as dielectric barriers for the DBD. The main difference with the earlier work~\cite{Pavan2022} is that the high voltage electrode is substituted by a glass slide of 1mm thickness with a thin ITO coating at the dry-side of the reactor, which is attached to the upper quartz plate. This electrode is 36mm long and centered over the point where the channel height is 3.4mm. High voltage is applied using a nanosecond pulse generator (Transient Plasma Systems SSPG-20X-HP1) at 8kHz and variable amplitude. The ground electrode is an aluminum plate. Electrical measurements are acquired using a high voltage probe (Lecroy PPE 20kV) and a Rogowski-coil style current monitor (Pearson model 6595) placed on the high voltage side of the circuit, close to the high-voltage probe. Signals are collected at 100ps resolution using a Teledyne Lecroy Waverunner 9254 oscilloscope. High speed video is taken using a high speed camera (Edgertronic SC2+), looking directly through the transparent high voltage electrode (along the z-axis in figure~\ref{fig:tapered_test_section}). For the images presented, the frame rate of the camera was set to 2kfps, as a compromise between accurate flame tracking under the electrode and isolation of individual pulses, and adequate resolution of the structures in low light. All signal synchronization is done using a delay generator (Berkeley Nucleonics Corp. Model 577-4C). 

\begin{figure*}[htb]
    \centering
\includegraphics[width=1\textwidth, trim={0cm 0.0cm 0cm 0cm}, clip]{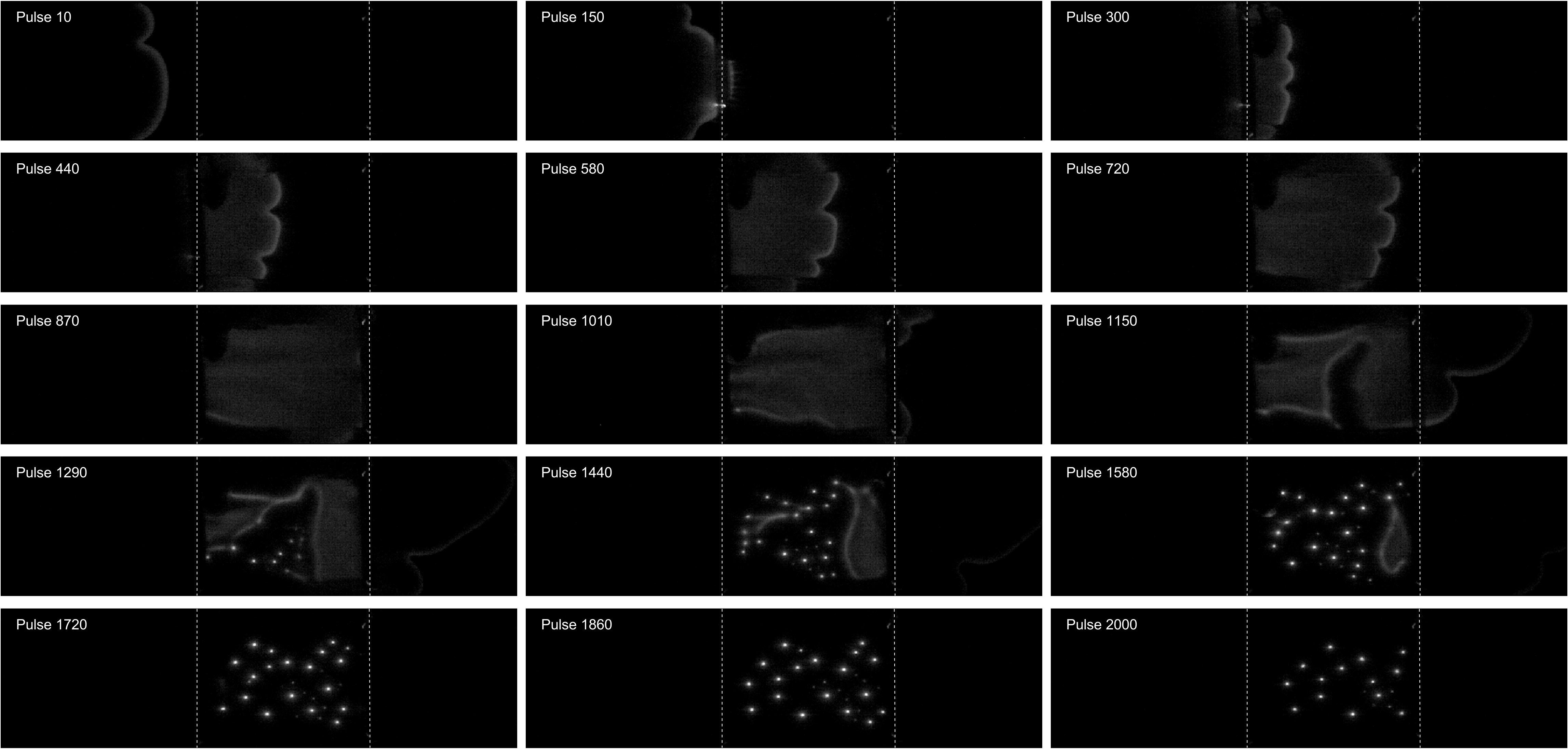}
    \caption{Interaction between premixed mesoscale flame propagating in narrow channel and NPD DBD plasma ignited by flame passage: in-situ strategy. The view presented is through the transparent ITO anode. Train of 2000 pulses applied at 20.4kV and 8kHz. Images correspond to visible-spectrum light accumulated in 4 pulses + flame emission (camera frame rate 2kfps). Electrode edge marked by dashed white lines.}
    \label{fig:regimes_flame_passage_overhead}
\end{figure*}

\begin{figure}[htb]
\includegraphics[width=0.4\textwidth, trim={0cm 0cm 0.5cm 0cm}, clip]{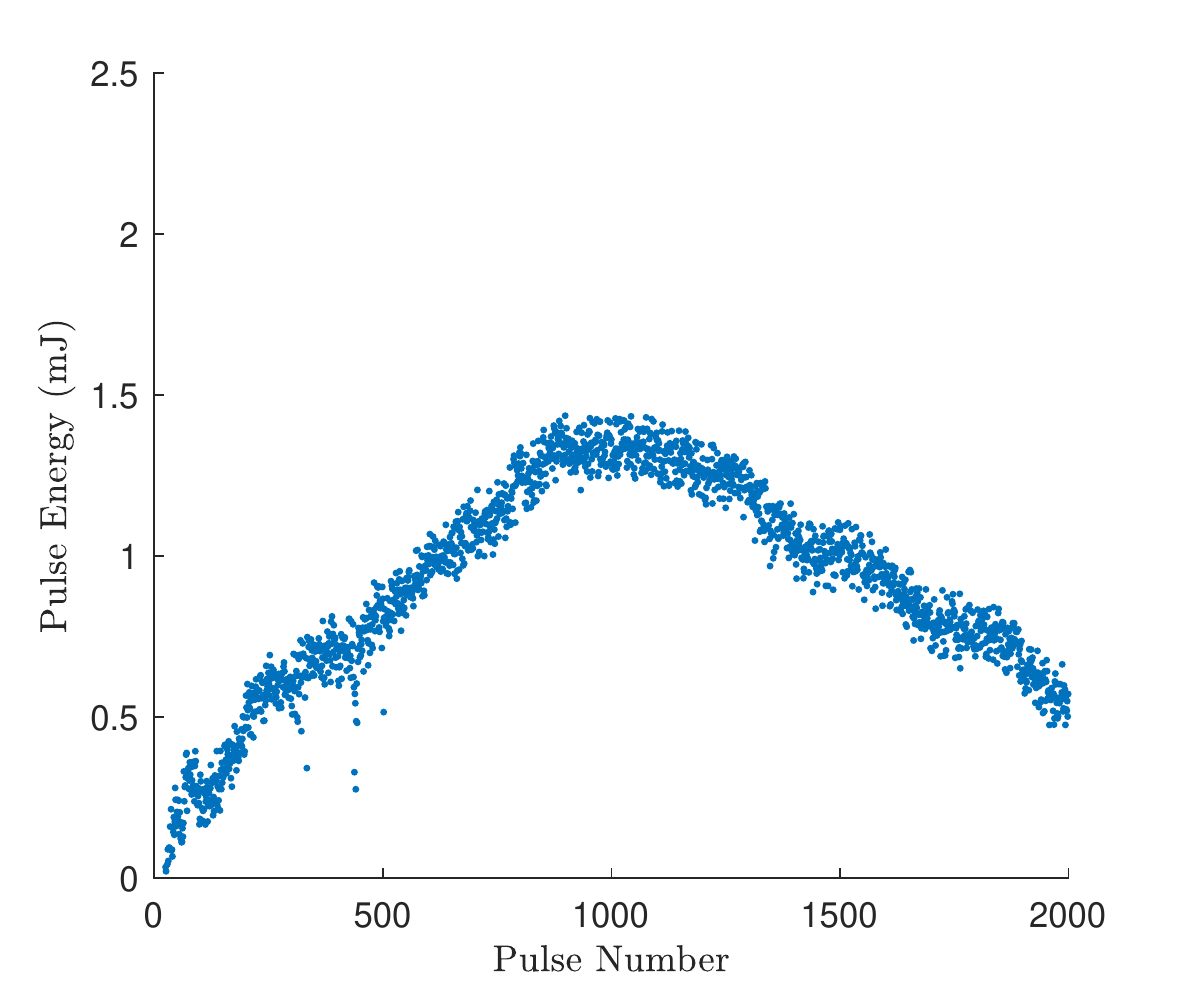}
    \caption{Per-pulse energy measurements corresponding to images shown in figure~\ref{fig:regimes_flame_passage_overhead}.}
    \label{fig:nrg_evo_overhead}
\end{figure}

Figure~\ref{fig:regimes_flame_passage_overhead} shows images of several pulses within a train of 2000 pulses, at 20.4kV, during passage of the laminar flame, in conditions phenomenologically similar to those depicted in section~\ref{section:transientflame}. The evolution of the energy deposition is pictured in figure~\ref{fig:nrg_evo_overhead}. For the \textbf{backward problem (flame-on-plasma)} the timeline is as follows: 
\begin{itemize}
    \item No visible discharge is observed for pulses applied before the flame front reaches the inter-electrode gap (pulse 10). 
    \item A uniform discharge ignites behind the flame as soon as its front reaches the electrode (pulse 150), in the region of burnt gas contained underneath the electrode (pulse 300). 
    \item The uniform discharge grows in size, as the flame propagates (pulse 440, 580, 720), until the flame reaches the end of the electrode (pulse 870). As the discharge grows in size, the per-pulse energy deposited increases. 
    \item The uniform discharge remains contained to the inter-electrode space, and detaches from the flame front as the flame continues its propagation beyond the electrode (pulse 1010, 1150). During this time the per-pulse energy has reached its maximum. This uniform region starts to shrink, in particular detaching from the burner walls that represent the coldest region of gas. 
    \item The discharge begins to transition to a microdischarge regime as the flame moves further away (pulse 1290, 1440) and eventually quenches (pulse 1580). As the discharge transitions to the filamentary mode and the discharge volume starts to shrink, the energy deposition starts to decrease. The filaments cover more of the volume, until the uniform contribution fully disappears (pulse 1720). 
    \item The microdischarges persist (pulse 1860, 2000) until the end of the train, even though the flame is no longer present. During this time the energy deposition decreases, likely due to the cooling gas conditions.
\end{itemize}

As the discharge and the flame are of comparable size, the two-way interaction is amplified and the plasma greatly modifies the flame properties in a dynamic manner. For the \textbf{forward problem (plasma-on-flame)} the timeline is as follows: 
\begin{itemize}
    \item The flame propagates in the channel with close to constant speed, comparable to that expected for the mixture at hand in a channel with heat loss. The flame front presents a two-lobe structure as a consequence of the finite dimension of the channel and the development of fluid instabilities. The shape is typical of flames in confined channels, and the structure is sometimes referred to as a tulip flame~\cite{Clanet1996, Gonzalez1992}.
    \item The flame presents more wrinkling (three-lobe structure instead of two) and speeds up (see below), as it enters the inter-electrode space (pulse 150, 300, 440, 580, 720).
    \item The flame seems to come to a halt as it reaches the end of the electrode (pulse 870, 1010).  
    \item The flame continues its propagation beyond the electrode, with a two-lobe structure (pulse 1150, 1290).
    \item As the flame detaches from the electrode and eventually quenches (pulse 1580), the flowfield structure of the post-combustion gases can be inferred from the visible microdischarges and uniform discharge patches.
\end{itemize}

\begin{figure*}[htb]
    \centering
\includegraphics[width=1\textwidth, trim={0cm 0.0cm 0cm 0cm}, clip]{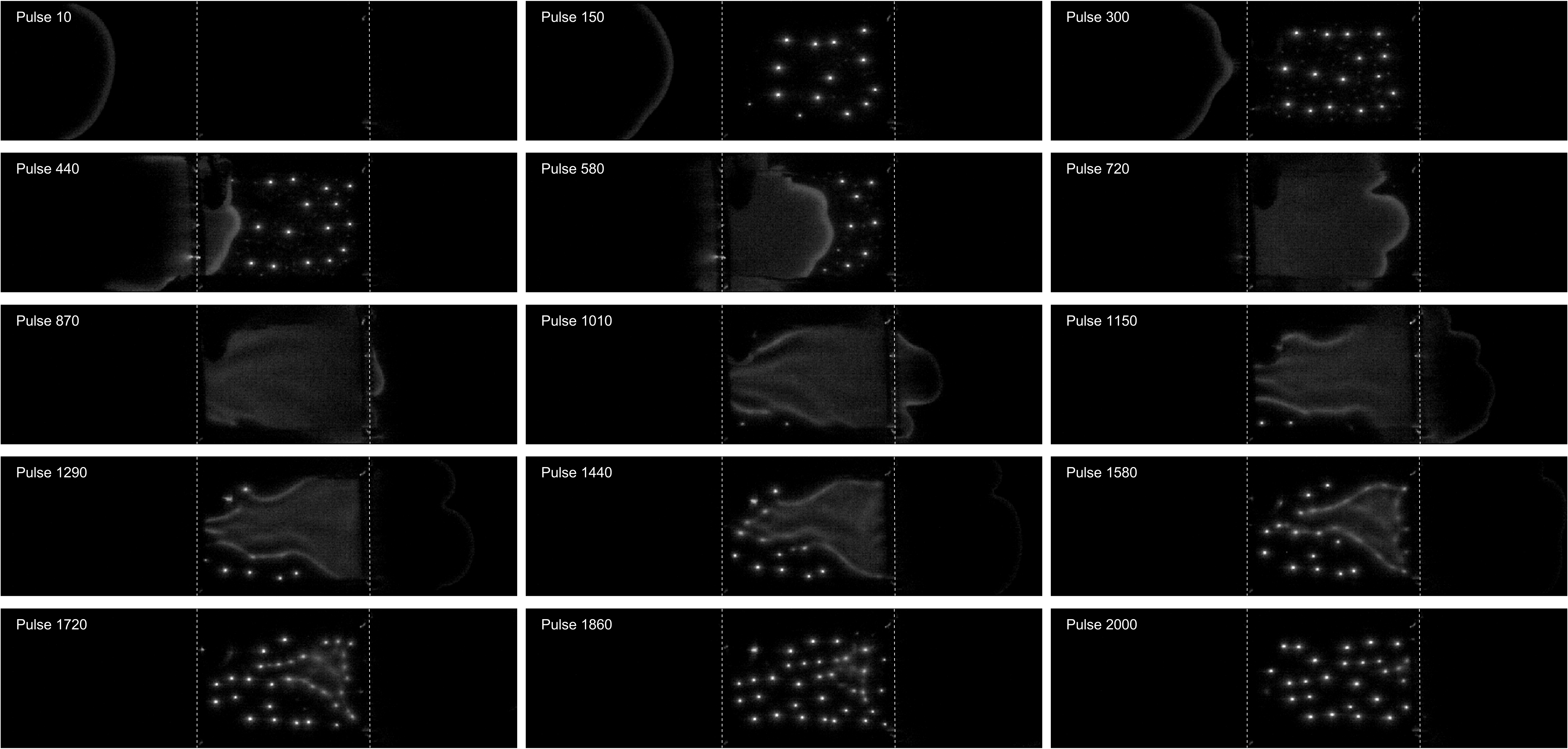}
    \caption{Interaction between premixed mesoscale flame propagating in narrow channel and NPD DBD plasma ignited by flame passage: ahead-of flame strategy. The view presented is through the transparent ITO anode. Train of 3200 pulses applied at 22.5kV and 8kHz. Images correspond to visible-spectrum light accumulated in 4 pulses + flame emission (camera frame rate 2kfps). Electrode edge marked by dashed white lines.}
\label{fig:regimes_flame_passage_overhead2}
\end{figure*}

\begin{figure}[htb]
         \centering
\includegraphics[width=0.4\textwidth, trim={0cm 0cm 0.5cm 0cm}, clip]{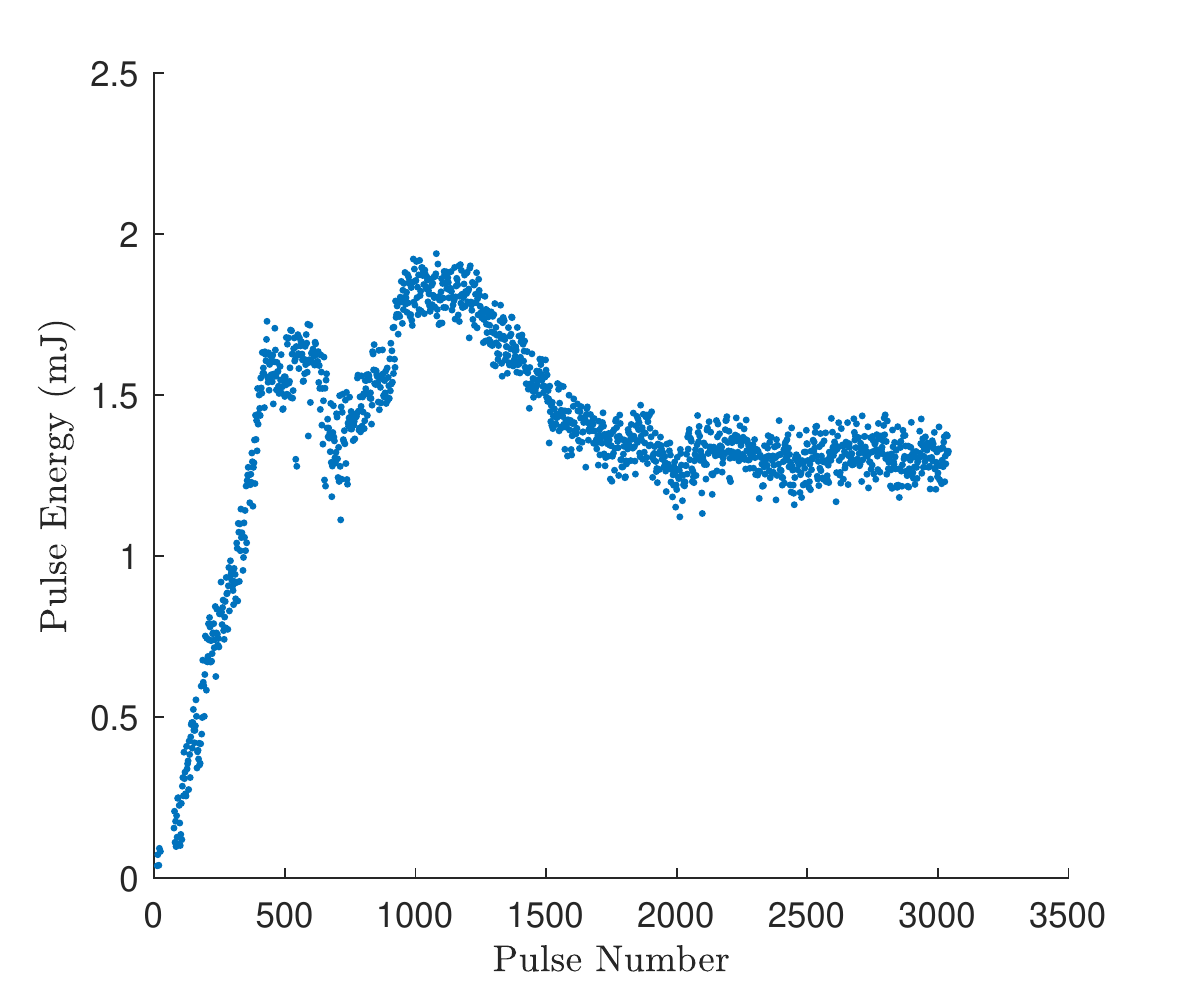}
    \caption{Per-pulse energy measurements corresponding to images shown in figure~\ref{fig:regimes_flame_passage_overhead2}.}
    \label{fig:nrg_evo_overhead2}
\end{figure}

\begin{figure}[htb]
         \centering
\includegraphics[width=0.5\textwidth, trim={3cm 0cm 2cm 0cm}, clip]{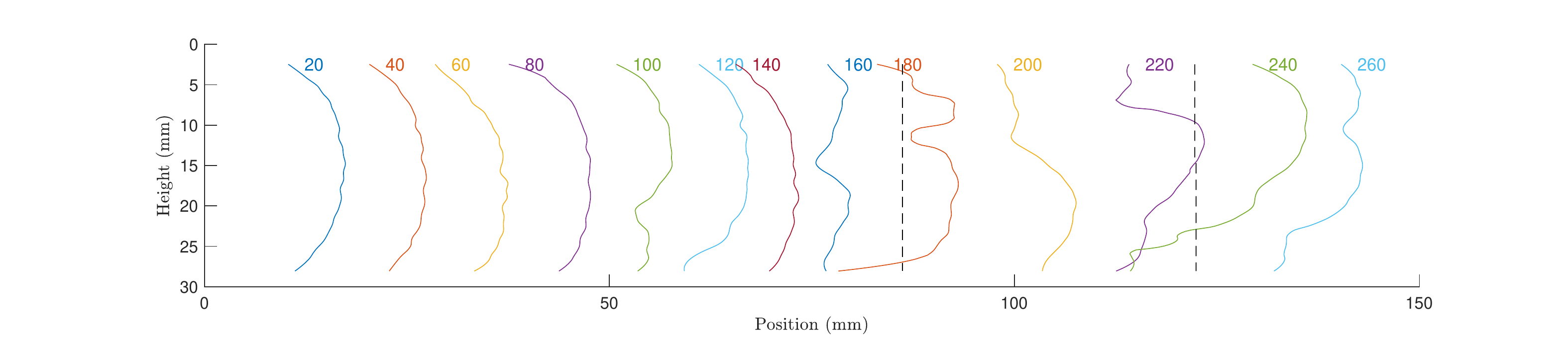}
    \caption{Shape and position of flame front at equally spaced time points. Dashed vertical lines mark the edges of the electrode and labels indicate the time in ms. Train of 2000 pulses applied at 22.5kV and 8kHz. Pulse train begins at approximately 120ms. Note that while the discharge conditions are the same as figure~\ref{fig:regimes_flame_passage_overhead2}, it is a different test, hence the slightly different flame structure.}
    \label{fig:flame_pos}
\end{figure}

A case with slightly higher voltage, e.g. 22.5kV, is presented in figures~\ref{fig:regimes_flame_passage_overhead2} and \ref{fig:nrg_evo_overhead2}. The chronology of the interaction is similar to the previous example, with the difference that in this case the applied voltage is high enough for breakdown to occur ahead of the flame, in the narrowing section of the channel, before the flame arrives to the electrodes. This strategy allows for pre-activation of the reactants, ahead of the flame, as compared to the in-situ strategy (fully coupled flame and plasma) described in the previous case. A few additional observations follow. From the discharge perspective, the discharge in the cold reactants (pulse 150, 300) proceeds in the microdischarge regime. As soon as the flame enters the inter-electrode space transition to uniform mode is observed in the burnt gas region (pulse 440, 580). As the flame quenches and the full inter-electrode gap sustains mostly micro-discharges (pulse 1720, 1860, 2000) the per-pulse energy remains constant and does not drop like in the previous case as the voltage is enough to sustain the discharge even in the cold gas. From the flame perspective, the enhanced flame wrinkling is observed before the flame reaches the electrode (pulse 300). The flame speed-up is also more noticeable in this higher voltage case. This flame speed increase, while the flame is underneath the electrodes, can be appreciated in figure~\ref{fig:flame_pos} by plotting the flame contours every 20ms: the flame clearly moves a greater distance, within 20ms, while located underneath the electrode as compared to other locations.   

In this configuration, the discharge transitions between uniform and filamentary regimes multiple times. If the applied voltage is high enough to allow breakdown in the cold reactant gas, the discharge often starts uniform for the first several pulses before transitioning to filaments. This is thought to be caused in part by a plasma thermal-chemical instability~\cite{Zhong2021_PSST,Rousso2020} and part due to slow gas flow since it was possible to achieve a uniform discharge in pure air with high flow rates but not in methane-air mixtures. Behind the flame front, the fuel oxidation reactions have largely been completed, the gas has been accelerated, and the temperature is high. All these aid in creating a uniform regime; since most oxidative reactions have ceased, the plasma thermal-chemical instability is less important, the fast moving gas prevents instability amplification (see section~\ref{sec:flow_fields}), and the higher temperature aids with diffusion, a uniformizing process. As the gas cools and slows, notably from the walls far behind the flame, the latter two effects are lessened and the discharge transitions back to filamentary.

\section{Implications for Enhancement and Control of Combustion by Plasma}
\label{sec:applications}
While the forward problem represents the driver for research in plasma-assisted combustion, the backward problem might well be the enabler for the success of the strategy. Several studies have focused on NPD plasmas in pin-to-pin configurations and steady environments, including air and post-combustion gases \cite{Kruger2002,Pancheshnyi2006,Rusterholtz2013UltrafastDischarges,Pai2009,Minesi2022,Xu2011,Xu2014ThermalAir}, and have set the foundation of our understanding of the plasma-chemical activation as well as the energy share that goes to other processes, such as shock wave generation or rapid gas heating. The implications of moving to higher pressure environments, engine-relevant fuels, or fast fluid flows, is still to be addressed. Whereas pin-to-pin discharges represent an ideal platform for optical studies, due to the fixed location of the plasma in space, and are already relevant for applications such as bluff-body stabilized burner configurations (section~\ref{sec:industrial}), other electrode configurations also need to be characterized. The diagnostics for situations where filamentary plasmas are left to move freely with the flow is more difficult to realize. In addition, and as motivated in previous sections, non-traditional studies of electrical breakdown, that move beyond the uniform and static gas background conditions, are also required to understand the development of NPD discharges in flames. 

Since this nonuniform and unsteady environment will guide the plasma dynamics and energy deposition process, it needs to be accounted for when designing an actuation strategy. In particular, our ability to address the grand challenges in combustion discussed in section~\ref{sec:intro} will require consideration of:
\begin{itemize}
    \item Plasma response to combustion environment, when designing actuation strategies for control of combustion dynamics. The plasma can evolve at the timescale of the instability, in the worst case resulting in a weakened discharge as the flame detaches from the flame-holder element. Our flameholding ability relies on accounting for this two-way coupling and controlling the plasma parameters by other means, if necessary. 
    \item Reinforcement of plasma in hot spots, and flamelets, when considering turbulent flames. Typically, the discharge volume is significantly smaller than the flame volume, and the location of the plasma source affects the degree of combustion enhancement, or disturbance. For situations where the discharge characteristic size is comparable to the flame features, reinforcement of the plasma in high temperature regions will be encountered. This can have a beneficial effect, if plasma deposition occurs close to the reaction front, but can also lead to wasted energy, if the energy deposition happens in the post-combustion gases.
    \item Realistic flow conditions, when characterizing plasma sources for practical system. The implications of fast flowing gases are typically overlooked and affect the plasma dynamics, the spatial structure of the energy deposition, the synergy between pulses, and the plasma regime transitions.
\end{itemize}

\section{Conclusions}
\label{sec:conclusions}
The problem of plasma-assisted combustion has been presented in terms of a two-way problem. The forward problem refers to the ability of impacting ignition and flames using a plasma. This side of the interaction is the driver of research in the field and is widely documented in the literature, both in terms of fundamental efforts and practical demonstrations. The backward problem refers to the influence of combustion on electrical breakdown and discharge dynamics. This side of the interaction is the focus of this paper and has been less documented in the literature, with most prior efforts focusing on detailed kinetic characterization of the plasma in uniform background gases. This article has presented a series of experiments that exemplify the relevance of addressing the backward problem, in particular when considering practical applications and dynamic combustion environments. It is argued that, when using pulsed nanosecond discharges for control of dynamic processes, or when the characteristic length scales of combustion are comparable to the inter-electrode gap length, the coupled interaction can not be overlooked. In these cases, the success of the strategy may well depend on properly accounting for the backward problem. 

\section*{Acknowledgements}
This work was partially funded by the Office of Naval Research (ONR) Young Investigator Award \emph{Transient Corona Discharges for Ignition and Flameholding in Navy Relevant Combustion}, N00014-21-1-2571. The authors would like to acknowledge fruitful collaborations throughout the years with the co-authors of some of the works reviewed in this paper: M. Martinez-Sanchez, A. Starikovskiy, R. Miles, F. Gomez del Campo, D. Weibel, S. Shanbhogue, and A. Ghoniem.



 \bibliographystyle{elsarticle-num} 
 \bibliography{references,extra_refs}

\begin{thebibliography}{100}
\expandafter\ifx\csname url\endcsname\relax
  \def\url#1{\texttt{#1}}\fi
\expandafter\ifx\csname urlprefix\endcsname\relax\def\urlprefix{URL }\fi
\expandafter\ifx\csname href\endcsname\relax
  \def\href#1#2{#2} \def\path#1{#1}\fi

\bibitem{Heiser1994}
W.~Heiser, D.~Pratt, D.~Daley, U.~Mehta, {Hypersonic Airbreathing Propulsion},
  American Institute of Aeronautics and Astronautics, Inc., Washington, DC,
  1994.
\newblock \href {https://doi.org/10.2514/4.470356}
  {\path{doi:10.2514/4.470356}}.

\bibitem{Liu2020}
Q.~Liu, D.~Baccarella, T.~Lee, {Review of combustion stabilization for
  hypersonic airbreathing propulsion}, Progress in Aerospace Sciences 119
  (2020) 100636.
\newblock \href {https://doi.org/10.1016/j.paerosci.2020.100636}
  {\path{doi:10.1016/j.paerosci.2020.100636}}.

\bibitem{Gann2015}
B.~McGann, C.~D. Carter, T.~Ombrello, H.~Do, {Direct spectrum matching of
  laser-induced breakdown for concentration and gas density measurements in
  turbulent reacting flows}, Combustion and Flame 162~(12) (2015) 4479--4485.
\newblock \href {https://doi.org/10.1016/j.combustflame.2015.08.021}
  {\path{doi:10.1016/j.combustflame.2015.08.021}}.

\bibitem{Leonov2009}
S.~Leonov, D.~Yarantsev, C.~Carter, {Experiments on Electrically Controlled
  Flameholding on a Plane Wall in a Supersonic Airflow}, Journal of Propulsion
  and Power 25~(2) (2009) 289--294.
\newblock \href {https://doi.org/10.2514/1.38002} {\path{doi:10.2514/1.38002}}.

\bibitem{Lee2020The2018}
D.~S. Lee, D.~W. Fahey, A.~Skowron, M.~R. Allen, U.~Burkhardt, Q.~Chen, S.~J.
  Doherty, S.~Freeman, P.~M. Forster, J.~Fuglestvedt, A.~Gettelman, R.~R.
  De~Le{\'{o}}n, L.~L. Lim, M.~T. Lund, R.~J. Millar, B.~Owen, J.~E. Penner,
  G.~Pitari, M.~J. Prather, R.~Sausen, L.~J. Wilcox, {The contribution of
  global aviation to anthropogenic climate forcing for 2000 to 2018},
  Atmospheric Environment 244 (2020).
\newblock \href {https://doi.org/10.1016/j.atmosenv.2020.117834}
  {\path{doi:10.1016/j.atmosenv.2020.117834}}.

\bibitem{Turns2000}
S.~R. Turns, {An introduction to combustion, concepts and applications}, 2nd
  Edition, McGraw-Hill Higher Education, 2000.

\bibitem{Lieuwen2012}
T.~C. Lieuwen, {Unsteady combustor physics}, Cambridge University Press, 2012.
\newblock \href {https://doi.org/10.1017/9781108889001}
  {\path{doi:10.1017/9781108889001}}.

\bibitem{Masri2021}
A.~R. Masri, {Challenges for turbulent combustion}, Proceedings of the
  Combustion Institute 38~(1) (2021) 121--155.
\newblock \href {https://doi.org/10.1016/j.proci.2020.07.144}
  {\path{doi:10.1016/j.proci.2020.07.144}}.

\bibitem{Kobayashi2019ScienceCombustion}
H.~Kobayashi, A.~Hayakawa, K.~D.~A. Somarathne, E.~C. Okafor, {Science and
  technology of ammonia combustion}, Proceedings of the Combustion Institute
  37~(1) (2019) 109--133.
\newblock \href {https://doi.org/10.1016/j.proci.2018.09.029}
  {\path{doi:10.1016/j.proci.2018.09.029}}.

\bibitem{Ju2015PlasmaChemistry}
Y.~Ju, W.~Sun, {Plasma assisted combustion: Dynamics and chemistry}, Progress
  in Energy and Combustion Science 48 (2015) 21--83.
\newblock \href {https://doi.org/10.1016/j.pecs.2014.12.002}
  {\path{doi:10.1016/j.pecs.2014.12.002}}.

\bibitem{Starikovsky2015}
A.~Starikovskiy, {Physics and chemistry of plasma-assisted combustion},
  Philosophical Transactions of the Royal Society A: Mathematical, Physical and
  Engineering Sciences 373 (2015) 20150074.
\newblock \href {https://doi.org/10.1098/rsta.2015.0074}
  {\path{doi:10.1098/rsta.2015.0074}}.

\bibitem{Adamovich2015ChallengesCombustion}
I.~V. Adamovich, W.~R. Lempert, {Challenges in understanding and predictive
  model development of plasma-assisted combustion}, Plasma Physics and
  Controlled Fusion 57~(1) (2015).
\newblock \href {https://doi.org/10.1088/0741-3335/57/1/014001}
  {\path{doi:10.1088/0741-3335/57/1/014001}}.

\bibitem{Shinohara2009}
K.~Shinohara, N.~Takada, K.~Sasaki, {Enhancement of burning velocity in
  premixed burner flame by irradiating microwave power}, Journal of Physics D:
  Applied Physics 42~(18) (2009) 182008.

\bibitem{Stockman2009}
E.~S. Stockman, S.~H. Zaidi, R.~B. Miles, C.~D. Carter, M.~D. Ryan,
  {Measurements of combustion properties in a microwave enhanced flame},
  Combustion and Flame 156~(7) (2009) 1453--1461.

\bibitem{Yu2010}
X.~Yu, J.~Peng, P.~Yang, R.~Sun, Y.~Yi, Y.~Zhao, D.~Chen, J.~Yu, {Enhancement
  of a laminar premixed methane/oxygen/nitrogen flame speed using
  femtosecond-laser-induced plasma}, Applied Physics Letters 97~(1) (2010)
  11503.

\bibitem{Elkholy2018BurningPlatform}
A.~Elkholy, Y.~Shoshyn, S.~Nijdam, J.~A. van Oijen, E.~M. van Veldhuizen,
  U.~Ebert, L.~P. de~Goey, {Burning velocity measurement of lean methane-air
  flames in a new nanosecond DBD microplasma burner platform}, Experimental
  Thermal and Fluid Science 95 (2018) 18--26.
\newblock \href {https://doi.org/10.1016/j.expthermflusci.2018.01.011}
  {\path{doi:10.1016/j.expthermflusci.2018.01.011}}.

\bibitem{Ombrello2010_1}
T.~Ombrello, S.~H. Won, Y.~Ju, S.~Williams, {Flame propagation enhancement by
  plasma excitation of oxygen. Part II: effects of O2 (a1{$\Delta$}g)},
  Combustion and Flame 157~(10) (2010) 1906--1915.

\bibitem{Pilla2006}
G.~Pilla, D.~Galley, D.~A. Lacoste, F.~Lacas, D.~Veynante, C.~O. Laux,
  {Stabilization of a Turbulent premixed flame using a nanosecond repetitively
  pulsed discharge}, IEEE Transactions on Plasma Science 34~(6) (2006)
  2471--2477.

\bibitem{Pham2011StabilizationDischarges}
Q.~L.~L. Pham, D.~A. Lacoste, C.~O. Laux, {Stabilization of a premixed
  methane-air flame using nanosecond repetitively pulsed discharges}, IEEE
  Transactions on Plasma Science 39~(11) (2011) 2264--2265.
\newblock \href {https://doi.org/10.1109/TPS.2011.2163806}
  {\path{doi:10.1109/TPS.2011.2163806}}.

\bibitem{DiSabatino2020EnhancementDischarges}
F.~Di~Sabatino, D.~A. Lacoste, {Enhancement of the lean stability and blow-off
  limits of methane-air swirl flames at elevated pressures by nanosecond
  repetitively pulsed discharges}, Journal of Physics D: Applied Physics
  53~(35) (2020) 355201.
\newblock \href {https://doi.org/10.1088/1361-6463/ab8f54}
  {\path{doi:10.1088/1361-6463/ab8f54}}.

\bibitem{Barbosa2015InfluenceCombustor}
S.~Barbosa, G.~Pilla, D.~A. Lacoste, P.~Scouflaire, S.~Ducruix, C.~O. Laux,
  D.~Veynante, {Influence of nanosecond repetitively pulsed discharges on the
  stability of a swirled propane/air burner representative of an aeronautical
  combustor}, Philosophical Transactions of the Royal Society A: Mathematical,
  Physical and Engineering Sciences 373~(2048) (2015).
\newblock \href {https://doi.org/10.1098/rsta.2014.0335}
  {\path{doi:10.1098/rsta.2014.0335}}.

\bibitem{Choe2018BlowoffFlow}
J.~Choe, W.~Sun, {Blowoff hysteresis, flame morphology and the effect of plasma
  in a swirling flow}, Journal of Physics D: Applied Physics 51~(36) (2018).
\newblock \href {https://doi.org/10.1088/1361-6463/aad4dc}
  {\path{doi:10.1088/1361-6463/aad4dc}}.

\bibitem{Moeck2013}
J.~P. Moeck, D.~A. Lacoste, C.~O. Laux, C.~O. Paschereit, {Control of
  combustion dynamics in a swirl-stabilized combustor with nanosecond
  repetitively pulsed discharges}, 51st AIAA Meeting, Grapevine, Texas (2013)
  2013--0565.

\bibitem{Shanbhogue2022}
S.~J. Shanbhogue, C.~A. Pavan, D.~E. Weibel, F.~Gomez~del Campo,
  C.~Guerra-Garcia, A.~F. Ghoniem, {Control of Large-Amplitude Combustion
  Oscillations Using Nanosecond Repetitively Pulsed Plasmas}, Journal of
  Propulsion and Power (2023) 1--13\href {https://doi.org/10.2514/1.B38883}
  {\path{doi:10.2514/1.B38883}}.

\bibitem{Faingold2021APlasma}
G.~Faingold, J.~K. Lefkowitz, {A numerical investigation of NH3/O2/He ignition
  limits in a non-thermal plasma}, Proceedings of the Combustion Institute
  38~(4) (2021) 6661--6669.
\newblock \href {https://doi.org/10.1016/j.proci.2020.08.033}
  {\path{doi:10.1016/j.proci.2020.08.033}}.

\bibitem{Shioyoke2018NumericalFlame}
A.~Shioyoke, J.~Hayashi, R.~Murai, N.~Nakatsuka, F.~Akamatsu, {Numerical
  Investigation on Effects of Nonequilibrium Plasma on Laminar Burning Velocity
  of Ammonia Flame}, Energy and Fuels 32~(3) (2018) 3824--3832.
\newblock \href {https://doi.org/10.1021/acs.energyfuels.7b02733}
  {\path{doi:10.1021/acs.energyfuels.7b02733}}.

\bibitem{Taneja2021}
T.~S. Taneja, S.~Yang, {Numerical modeling of plasma assisted pyrolysis and
  combustion of ammonia}, in: AIAA Scitech 2021 Forum, no. January, 2021, pp.
  2021--1972.
\newblock \href {https://doi.org/10.2514/6.2021-1972}
  {\path{doi:10.2514/6.2021-1972}}.

\bibitem{Choe2021PlasmaEnhancement}
J.~Choe, W.~Sun, T.~Ombrello, C.~Carter, {Plasma assisted ammonia combustion:
  Simultaneous NOx reduction and flame enhancement}, Combustion and Flame 228
  (2021) 430--432.
\newblock \href {https://doi.org/10.1016/j.combustflame.2021.02.016}
  {\path{doi:10.1016/j.combustflame.2021.02.016}}.

\bibitem{Popov2022RelaxationStabilization}
N.~A. Popov, S.~M. Starikovskaia, {Relaxation of electronic excitation in
  nitrogen/oxygen and fuel/air mixtures: fast gas heating in plasma-assisted
  ignition and flame stabilization}, Progress in Energy and Combustion Science
  91 (2022) 100928.
\newblock \href {https://doi.org/10.1016/j.pecs.2021.100928}
  {\path{doi:10.1016/j.pecs.2021.100928}}.

\bibitem{Guerra-Garcia2015}
C.~Guerra-Garcia, M.~Martinez-Sanchez, {Counterflow nonpremixed flame DC
  displacement under AC electric field}, Combustion and Flame 162~(11) (2015)
  4254--4263.
\newblock \href {https://doi.org/10.1016/j.combustflame.2015.07.038}
  {\path{doi:10.1016/j.combustflame.2015.07.038}}.

\bibitem{Xu2011}
D.~A. Xu, D.~A. Lacoste, D.~L. Rusterholtz, P.~Q. Elias, G.~D. Stancu, C.~O.
  Laux, {Experimental study of the hydrodynamic expansion following a
  nanosecond repetitively pulsed discharge in air}, Applied Physics Letters
  99~(12) (2011) 121502.
\newblock \href {https://doi.org/10.1063/1.3641413}
  {\path{doi:10.1063/1.3641413}}.

\bibitem{Raizer1991}
Y.~Raitzer, {Gas Discharge Physics}, Springer-Verlag, Berlin, 1991.

\bibitem{Starikovskiy2013}
A.~Starikovskiy, N.~Aleksandrov, {Review: Plasma-assisted ignition and
  combustion}, Progress in Energy and Combustion Science 39~(1) (2013) 61--110.
\newblock \href {https://doi.org/10.1016/j.pecs.2012.05.003}
  {\path{doi:10.1016/j.pecs.2012.05.003}}.

\bibitem{Nagaraja2013Multi-scaleGeometry}
S.~Nagaraja, V.~Yang, I.~Adamovich, {Multi-scale modelling of pulsed nanosecond
  dielectric barrier plasma discharges in plane-to-plane geometry}, Journal of
  Physics D: Applied Physics 46~(15) (2013).
\newblock \href {https://doi.org/10.1088/0022-3727/46/15/155205}
  {\path{doi:10.1088/0022-3727/46/15/155205}}.

\bibitem{Popov2007}
N.~A. Popov, {The effect of nonequilibrium excitation on the ignition of
  hydrogen-oxygen mixtures}, High Temperature 45~(2) (2007) 261--279.
\newblock \href {https://doi.org/10.1134/S0018151X07020174}
  {\path{doi:10.1134/S0018151X07020174}}.

\bibitem{Aleksandrov2010}
N.~L. Aleksandrov, S.~V. Kindysheva, M.~M. Nudnova, A.~Y. Starikovskiy,
  {Mechanism of ultra-fast heating in a non-equilibrium weakly ionized air
  discharge plasma in high electric fields}, Journal of Physics D: Applied
  Physics 43~(25) (2010) 255201.
\newblock \href {https://doi.org/10.1088/0022-3727/43/25/255201}
  {\path{doi:10.1088/0022-3727/43/25/255201}}.

\bibitem{Starikovskiy2012Plasma-assistedTransition}
A.~Starikovskiy, N.~Aleksandrov, A.~Rakitin, {Plasma-assisted ignition and
  deflagration-to-detonation transition}, Philosophical Transactions of the
  Royal Society A: Mathematical, Physical and Engineering Sciences 370~(1960)
  (2012) 740--773.
\newblock \href {https://doi.org/10.1098/rsta.2011.0344}
  {\path{doi:10.1098/rsta.2011.0344}}.

\bibitem{Hagelaar2005}
G.~J.~M. Hagelaar, L.~C. Pitchford, {Solving the Boltzmann equation to obtain
  electron transport coefficients and rate coefficients for fluid models},
  Plasma Sources Science and Technology 14~(4) (2005) 722--733.
\newblock \href {https://doi.org/10.1088/0963-0252/14/4/011}
  {\path{doi:10.1088/0963-0252/14/4/011}}.

\bibitem{lxcat_hayashi}
M.~Hayashi, \href{www.lxcat.net/Hayashi}{Hayashi database} (retrieved on
  November 9, 2021).
\newline\urlprefix\url{www.lxcat.net/Hayashi}

\bibitem{lxcat_phelps}
A.~Phelps, \href{www.lxcat.net/Phelps}{Phelps database} (retrieved on November
  9, 2021).
\newline\urlprefix\url{www.lxcat.net/Phelps}

\bibitem{lxcat_siglo}
A.~Phelps, L.~Pitchford, \href{www.lxcat.net/SIGLO}{Siglo database} (retrieved
  on November 9, 2021).
\newline\urlprefix\url{www.lxcat.net/SIGLO}

\bibitem{Dhali1987}
S.~K. Dhali, P.~F. Williams, {Two-dimensional studies of streamers in gases},
  Journal of Applied Physics 62~(12) (1987) 4696--4707.

\bibitem{Hagelaar2000}
G.~J.~M. Hagelaar, G.~M.~W. Kroesen, G.~J.~M. Hageelar, G.~M.~W. Kroesen,
  {Speeding up fluid models for gas discharges by implicit treatment of the
  electron energy source term}, Journal of Computational Physics 159~(1) (2000)
  1--12.

\bibitem{Pancheshnyi2008}
S.~Pancheshnyi, P.~Segur, J.~Capeillere, A.~Bourdon, {Numerical simulation of
  filamentary discharges with parallel adaptive mesh refinement}, Journal of
  Computational Physics 227~(13) (2008) 6574--6590.

\bibitem{Lacoste2022}
D.~A. Lacoste, {Flames with plasmas}, Proceedings of the Combustion Institute
  (2022).
\newblock \href {https://doi.org/10.1016/j.proci.2022.06.025}
  {\path{doi:10.1016/j.proci.2022.06.025}}.

\bibitem{Sun2011_2}
W.~Sun, M.~Uddi, T.~Ombrello, S.~H. Won, C.~Carter, Y.~Ju, {Effects of
  non-equilibrium plasma discharge on counterflow diffusion flame extinction},
  Proceedings of the Combustion Institute 33~(2) (2011) 3211--3218.

\bibitem{Sun2013}
W.~Sun, S.~H. Won, T.~Ombrello, C.~Carter, Y.~Ju, {Direct ignition and S-curve
  transition by in situ nano-second pulsed discharge in methane/oxygen/helium
  counterflow flame}, Proceedings of the Combustion Institute 34~(1) (2013)
  847--855.

\bibitem{Stange2005}
S.~Stange, Y.~Kim, V.~Ferreri, {L.A.Rosocha}, D.~M. Coates, {Flame Images
  Indicating Combustion Enhancement by Dielectric Barrier Discharges}, IEEE
  Transactions on Plasma Science 33~(2) (2005).

\bibitem{Pai2010}
D.~Pai, D.~A. Lacoste, C.~O. Laux, {Transitions between corona, glow, and spark
  regimes of nanosecond repetitively pulsed discharges in air at atmospheric
  pressure}, Journal of Applied Physics 107~(9) (2010) 93303.

\bibitem{Kim2019Plasma-AssistedConditions}
W.~Kim, J.~Cohen, {Plasma-Assisted Combustor Dynamics Control at Realistic Gas
  Turbine Conditions}, Combustion Science and Technology (2019) 1--20\href
  {https://doi.org/10.1080/00102202.2019.1676743}
  {\path{doi:10.1080/00102202.2019.1676743}}.

\bibitem{Shanbhogue2022ActivePlasmas}
S.~Shanbhogue, D.~Weibel, F.~Gomez~del Campo, C.~Guerra-Garcia, A.~Ghoniem,
  {Active control of large amplitude combustion oscillations using nanosecond
  repetitively pulsed plasmas}, in: AIAA Scitech 2022 Forum, 2022, pp.
  2022--1450.
\newblock \href {https://doi.org/10.2514/6.2022-1450}
  {\path{doi:10.2514/6.2022-1450}}.

\bibitem{Mao2022_model}
X.~Mao, H.~Zhong, T.~Zhang, A.~Starikovskiy, Y.~Ju, {Modeling of the effects of
  non-equilibrium excitation and electrode geometry on H2/air ignition in a
  nanosecond plasma discharge}, Combustion and Flame 240 (2022) 112046.
\newblock \href {https://doi.org/10.1016/j.combustflame.2022.112046}
  {\path{doi:10.1016/j.combustflame.2022.112046}}.

\bibitem{Bane2015}
S.~P. Bane, J.~L. Ziegler, J.~E. Shepherd, {Investigation of the effect of
  electrode geometry on spark ignition}, Combustion and Flame 162~(2) (2015)
  462--469.
\newblock \href {https://doi.org/10.1016/j.combustflame.2014.07.017}
  {\path{doi:10.1016/j.combustflame.2014.07.017}}.

\bibitem{Boeuf2013}
J.~P. Boeuf, L.~L. Yang, L.~C. Pitchford, {Dynamics of a guided streamer
  (plasma bullet) in a helium jet in air at atmospheric pressure}, Journal of
  Physics D: Applied Physics 46~(1) (2013) 15201.

\bibitem{Breden2012_3}
D.~Breden, K.~Miki, L.~L. Raja, {Self-consistent two-dimensional modeling of
  cold atmospheric-pressure plasma jets/bullets}, Plasma Sources Science and
  Technology 21~(3) (2012) 34011.

\bibitem{Naidis2010}
G.~V. Naidis, {Modelling of streamer propagation in atmospheric-pressure helium
  plasma jets}, Journal of Physics D: Applied Physics 43~(40) (2010) 402001.

\bibitem{Naidis2012}
G.~V. Naidis, {Simulation of a Single Streamer Traveling along two
  counterpropagating Helium jets in ambient air}, IEEE Transactions on Plasma
  Science 40~(11) (2012) 2866--2869.

\bibitem{Douat2011}
C.~Douat, M.~Fleury, M.~Laroussi, V.~Puech, {Interactions Between Two
  Counterpropagating Plasma Bullets}, IEEE Transactions on Plasma Science
  39~(11) (2011) 2298--2299.

\bibitem{Naidis2011}
G.~V. Naidis, {Modelling of plasma bullet propagation along a helium jet in
  ambient air}, Journal of Physics D: Applied Physics 44~(21) (2011) 215203.

\bibitem{Breden2011}
D.~Breden, K.~Miki, L.~L. Raja, {Computational study of cold atmospheric
  nanosecond pulsed helium plasma jet in air}, Applied Physics Letters 99~(11)
  (2011) 111501.

\bibitem{Guerra-Garcia2015CanEffect}
C.~Guerra-Garcia, M.~Martinez-Sanchez, {Can insulating gaseous layers provide a
  dielectric barrier discharge effect?}, Applied Physics Letters 106~(4) (2015)
  4--7.
\newblock \href {https://doi.org/10.1063/1.4906417}
  {\path{doi:10.1063/1.4906417}}.

\bibitem{Luque2010}
A.~Luque, U.~Ebert, {Sprites in varying air density: charge conservation,
  glowing negative trails and changing velocity}, Geophysical Research Letters
  37 (2010) L06806.

\bibitem{Luque2012}
A.~Luque, U.~Ebert, {Density models for streamer discharges: beyond cylindrical
  symmetry and homogeneous media}, Journal of Computational Physics 231~(3)
  (2012) 904--918.

\bibitem{Opaits2010}
D.~F. Opaits, M.~N. Shneider, P.~J. Howard, R.~B. Miles, G.~M. Milikh, {Study
  of streamers in gradient density air: table top modeling of red sprites},
  Geophysical Research Letters 37 (2010) L14801.

\bibitem{Sigmond2004}
R.~S. Sigmond, T.~Sigmond, L.~Rolfseng, A.~F. Bohman, F.~T. Stormo,
  L.~Hvidsten, {The aiming of the bolt: how a flashover finds the weak spot},
  IEEE Transactions on Plasma Science 32~(5) (2004) 1812--1818.

\bibitem{Babaeva2009}
N.~Y. Babaeva, M.~J. Kushner, {Effect of inhomogeneities on streamer
  propagation: I. Intersection with isolated bubbles and particles}, Plasma
  Sources Science and Technology 18~(3) (2009) 35009.

\bibitem{Babaeva2009_3}
N.~Y. Babaeva, M.~J. Kushner, {Effect of inhomogeneities on streamer
  propagation: II. Streamer dynamics in high pressure humid air with bubbles},
  Plasma Sources Science and Technology 18~(3) (2009) 35010.

\bibitem{Pillai2022}
N.~Pillai, N.~L. Sponsel, J.~T. Mast, M.~J. Kushner, I.~A. Bolotnov,
  K.~Stapelmann, {Plasma breakdown in bubbles passing between two pin
  electrodes}, Journal of Physics D: Applied Physics 55~(47) (2022) 475203.
\newblock \href {https://doi.org/10.1088/1361-6463/ac9538}
  {\path{doi:10.1088/1361-6463/ac9538}}.

\bibitem{Babaeva2009_2}
N.~Y. Babaeva, M.~J. Kushner, {Structure of positive streamers inside gaseous
  bubbles immersed in liquids}, Journal of Physics D: Applied Physics 42~(13)
  (2009) 132003.
\newblock \href {https://doi.org/10.1088/0022-3727/42/13/132003}
  {\path{doi:10.1088/0022-3727/42/13/132003}}.

\bibitem{Bruggeman2009}
P.~Bruggeman, C.~Leys, {Non-thermal plasmas in and in contact with liquids},
  Journal of Physics D: Applied Physics 42~(5) (2009) 53001.

\bibitem{Starikovskiy2020}
A.~Y. Starikovskiy, N.~L. Aleksandrov, {Blocking streamer development by plane
  gaseous layers of various densities}, Plasma Sources Science and Technology
  29~(3) (2020) 034002.
\newblock \href {https://doi.org/10.1088/1361-6595/ab5837}
  {\path{doi:10.1088/1361-6595/ab5837}}.

\bibitem{Guerra2013}
C.~Guerra-Garcia, M.~Martinez-Sanchez, {Gas-confined barrier discharges: a
  simplified model for plasma dynamics in flame environments}, Journal of
  Physics D: Applied Physics 46~(34) (2013) 345204.
\newblock \href {https://doi.org/10.1088/0022-3727/46/34/345204}
  {\path{doi:10.1088/0022-3727/46/34/345204}}.

\bibitem{Guerra-Garcia2015a}
C.~Guerra-Garcia, M.~Martinez-Sanchez, R.~B. Miles, A.~Starikovskiy, {Localized
  pulsed nanosecond discharges in a counterflow nonpremixed flame environment},
  Plasma Sources Science and Technology 24~(5) (2015).
\newblock \href {https://doi.org/10.1088/0963-0252/24/5/055010}
  {\path{doi:10.1088/0963-0252/24/5/055010}}.

\bibitem{Govinda2006}
G.~G. Raju, {Gaseous electronics. Theory and practice}, Taylor and Francis
  Group, 2006.

\bibitem{Chantry1981}
P.~J. Chantry, R.~E. Wootton, {A critique of methods for calculating the
  dielectric strength of gas mixtures}, Journal of Applied Physics 52~(4)
  (1981) 2731--2739.

\bibitem{Maric2005}
D.~Maric, M.~Radmilovic-Radenovic, Z.~L. Petrovic, {On parametrization and
  mixture laws for electron ionization coefficients}, European Physical Journal
  D 35 (2005) 313--321.

\bibitem{Guerra-Garcia2015Non-thermalEnvironments}
C.~Guerra-Garcia, {Non-thermal plasmas in flames and other inhomogeneous
  environments}, Ph.D. thesis, Massachusetts Institute of Technology, PhD
  thesis (2015).

\bibitem{Adamovich2009}
I.~V. Adamovich, M.~Nishihara, I.~Choi, M.~Uddi, W.~R. Lempert, {Energy
  coupling to the plasma in repetitive nanosecond pulse discharges}, Physics of
  Plasmas 16~(11) (2009) 113505.
\newblock \href {https://doi.org/10.1063/1.3264740}
  {\path{doi:10.1063/1.3264740}}.

\bibitem{Gadkari2017}
S.~Gadkari, S.~Gu, {Numerical investigation of co-axial DBD: Influence of
  relative permittivity of the dielectric barrier, applied voltage amplitude,
  and frequency}, Physics of Plasmas 24~(5) (2017) 053517.
\newblock \href {https://doi.org/10.1063/1.4982657}
  {\path{doi:10.1063/1.4982657}}.

\bibitem{Rousso2020}
A.~C. Rousso, B.~M. Goldberg, T.~Y. Chen, S.~Wu, A.~Dogariu, R.~B. Miles,
  E.~Kolemen, Y.~Ju, {Time and space resolved diagnostics for plasma
  thermal-chemical instability of fuel oxidation in nanosecond plasma
  discharges}, Plasma Sources Science and Technology 29~(10) (2020) 105012.
\newblock \href {https://doi.org/10.1088/1361-6595/abb7be}
  {\path{doi:10.1088/1361-6595/abb7be}}.

\bibitem{Zhong2019}
H.~Zhong, M.~N. Shneider, M.~S. Mokrov, Y.~Ju, {Thermal-chemical instability of
  weakly ionized plasma in a reactive flow}, Journal of Physics D: Applied
  Physics 52~(48) (2019) 484001.
\newblock \href {https://doi.org/10.1088/1361-6463/ab3d69}
  {\path{doi:10.1088/1361-6463/ab3d69}}.

\bibitem{Leonov2016}
S.~B. Leonov, I.~V. Adamovich, V.~R. Soloviev, {Dynamics of near-surface
  electric discharges and mechanisms of their interaction with the airflow},
  Plasma Sources Science and Technology 25~(6) (2016) 063001.
\newblock \href {https://doi.org/10.1088/0963-0252/25/6/063001}
  {\path{doi:10.1088/0963-0252/25/6/063001}}.

\bibitem{Leonov2008}
S.~B. Leonov, D.~A. Yarantsev, {Near-Surface Electrical Discharge in Supersonic
  Airflow: Properties and Flow Control}, Journal of Propulsion and Power 24~(6)
  (2008) 1168--1181.
\newblock \href {https://doi.org/10.2514/1.24585} {\path{doi:10.2514/1.24585}}.

\bibitem{Leonov2014}
S.~B. Leonov, V.~Petrishchev, I.~V. Adamovich, {Dynamics of energy coupling and
  thermalization in barrier discharges over dielectric and weakly conducting
  surfaces on micro-s to ms time scales}, Journal of Physics D: Applied Physics
  47~(46) (2014) 465201.
\newblock \href {https://doi.org/10.1088/0022-3727/47/46/465201}
  {\path{doi:10.1088/0022-3727/47/46/465201}}.

\bibitem{Houpt2017}
A.~Houpt, B.~Hedlund, S.~Leonov, T.~Ombrello, C.~Carter, {Quasi-DC electrical
  discharge characterization in a supersonic flow}, Experiments in Fluids
  58~(4) (2017) 25.
\newblock \href {https://doi.org/10.1007/s00348-016-2295-5}
  {\path{doi:10.1007/s00348-016-2295-5}}.

\bibitem{Guerra2023_scitech}
C.~Guerra-Garcia, C.~A. Pavan, S.~Rao, R.~J. Dijoud, {Influence of Airflow on
  Nanosecond Pulsed Discharges}, in: AIAA SciTech 2023 Forum, American
  Institute of Aeronautics and Astronautics, Reston, Virginia, 2023, pp. 1--8.
\newblock \href {https://doi.org/10.2514/6.2023-1866}
  {\path{doi:10.2514/6.2023-1866}}.

\bibitem{Pavan2020InvestigationsModels}
C.~Pavan, M.~Martinez-Sanchez, C.~Guerra-Garcia, {Investigations of positive
  streamers as quasi-steady structures using reduced order models}, Plasma
  Sources Science and Technology 29~(9) (2020) 95004.
\newblock \href {https://doi.org/10.1088/1361-6595/aba863}
  {\path{doi:10.1088/1361-6595/aba863}}.

\bibitem{Lawton1963}
J.~Lawton, F.~J. Weinberg, {Maximum ion currents from flames and the maximum
  practical effects of applied electric fields}, Proceedings of the Royal
  Society of London. Series A 277~(1371) (1964) 468--497.

\bibitem{Lawton1967}
J.~Lawton, F.~J. Weinberg, {Electrical aspects of combustion}, Oxford
  University Press, 1969.

\bibitem{Semenov1970}
E.~S. Semenov, A.~S. Sokolik, {Thermal and chemical ionization in flames},
  Combustion, Explosion, and Shock Waves 6~(1) (1970) 37--48.

\bibitem{Pedersen1993}
T.~Pedersen, R.~C. Brown, {Simulation of electric field effects in premixed
  methane flames}, Combustion and Flame 94~(4) (1993) 433--448.

\bibitem{Starik2002}
A.~M. Starik, N.~S. Titova, {Kinetics of ion formation in the volumetric
  reaction of methane with air}, Combustion, Explosion and Shock waves 38~(3)
  (2002) 253--268.

\bibitem{Cool1984}
T.~A. Cool, P.~J.~H. Tjossem, {Direct observations of chemiionization in
  hydrocarbon flames enhanced by laser excited CH}, Chemical Physics Letters
  111~(1-2) (1984) 82--88.

\bibitem{McLatchy1979}
C.~S. MacLatchy, {Langmuir probe measurements of ion density in
  atmospheric-pressure air-propane flame}, Combustion and Flame 36 (1979)
  171--178.

\bibitem{Ju2004}
Y.~Ju, S.~O. Macheret, M.~N. Shneider, R.~B. Miles, {Numerical study of the
  effect of microwave discharge on the premixed methane-air flame}, 40th
  AIAA/ASME/SAE/ASEE Joint Propulsion Conference and Exhibit (2004) 2004--3707.

\bibitem{Calcote1962}
H.~F. Calcote, {Ion production and recombination in flames}, Symposium
  (International) on Combustion 8~(1) (1961) 184--199.

\bibitem{Lawton1968}
J.~Lawton, P.~J. Mayo, F.~J. Weinberg, {Electrical control of gas flows in
  combustion processes}, Proceedings of the Royal Society of London. Series A
  303~(1474) (1968) 275--298.

\bibitem{Sullivan2004}
D.~J. Sullivan, S.~H. Zaidi, S.~O. Macheret, Y.~Ju, R.~B. Miles, {Microwave
  techniques for the combustion enhancement of laminar flames}, 40th
  AIAA/ASME/SAE/ASEE Joint Propulsion Conference and Exhibit (2004) 2004--3713.

\bibitem{Stockman2009_phd}
E.~M. Stockman, {Microwave enhanced combustion of laminar hydrocarbon flame
  fronts}, Ph.D. thesis, Princeton University (2009).

\bibitem{Michael2012}
J.~B. Michael, {Localized Microwave Pulsed Plasmas for Ignition and Flame Front
  Enhancement}, Ph.D. thesis, Princeton University (2012).

\bibitem{Pai2009}
D.~Pai, G.~D. Stancu, D.~A. Lacoste, C.~O. Laux, {Nanosecond repetitively
  pulsed discharges in air at atmospheric pressure - the glow regime}, Plasma
  Sources Science and Technology 18~(4) (2009) 45030.

\bibitem{Kruger2002}
C.~H. Kruger, C.~O. Laux, L.~Yu, D.~M. Packan, L.~Pierrot, {Nonequilibrium
  discharges in air and nitrogen plasmas at atmospheric pressure}, Pure and
  Applied Chemistry 74~(3) (2002) 337--347.

\bibitem{Palmer1974}
A.~J. Palmer, {A physical model on the initiation of atmospheric-pressure glow
  discharges}, Applied Physics Letters 25~(3) (1974) 138--140.

\bibitem{Levatter1980}
J.~I. Levatter, S.~Lin, {Necessary conditions for the homogeneous formation of
  pulsed avalanche discharges at high gas pressures}, Journal of Applied
  Physics 51~(1) (1980) 210--222.

\bibitem{Huang2014}
B.-D. Huang, K.~Takashima, X.-M. Zhu, Y.-K. Pu, {The influence of the
  repetition rate on the nanosecond pulsed pin-to-pin microdischarges}, Journal
  of Physics D: Applied Physics 47~(42) (2014) 422003.
\newblock \href {https://doi.org/10.1088/0022-3727/47/42/422003}
  {\path{doi:10.1088/0022-3727/47/42/422003}}.

\bibitem{Huang2018}
B.-D. Huang, E.~Carbone, K.~Takashima, X.-M. Zhu, U.~Czarnetzki, Y.-K. Pu, {The
  effect of the pulse repetition rate on the fast ionization wave discharge},
  Journal of Physics D: Applied Physics 51~(22) (2018) 225202.
\newblock \href {https://doi.org/10.1088/1361-6463/aabf2d}
  {\path{doi:10.1088/1361-6463/aabf2d}}.

\bibitem{Minesi2022}
N.~Q. Minesi, V.~P. Blanchard, E.~Pannier, G.~D. Stancu, C.~O. Laux,
  {Plasma-assisted combustion with nanosecond discharges. I: Discharge effects
  characterization in the burnt gases of a lean flame}, Plasma Sources Science
  and Technology 31~(4) (2022) 045029.
\newblock \href {https://doi.org/10.1088/1361-6595/ac5cd4}
  {\path{doi:10.1088/1361-6595/ac5cd4}}.

\bibitem{Pancheshnyi2006}
S.~V. Pancheshnyi, D.~A. Lacoste, A.~Bourdon, C.~O. Laux, {Ignition of
  propane-air mixtures by a repetitively pulsed nanosecond discharge}, IEEE
  Transactions on Plasma Science 34~(6) (2006) 2478--2487.
\newblock \href {https://doi.org/10.1109/TPS.2006.876421}
  {\path{doi:10.1109/TPS.2006.876421}}.

\bibitem{Lefkowitz2017}
J.~K. Lefkowitz, T.~Ombrello, {An exploration of inter-pulse coupling in
  nanosecond pulsed high frequency discharge ignition}, Combustion and Flame
  180 (2017) 136--147.
\newblock \href {https://doi.org/10.1016/j.combustflame.2017.02.032}
  {\path{doi:10.1016/j.combustflame.2017.02.032}}.

\bibitem{Mao2022_interpulse}
X.~Mao, H.~Zhong, Z.~Wang, T.~Ombrello, Y.~Ju, {Effects of inter-pulse coupling
  on nanosecond pulsed high frequency discharge ignition in a flowing mixture},
  Proceedings of the Combustion Institute (2022) 1--8\href
  {https://doi.org/10.1016/j.proci.2022.06.018}
  {\path{doi:10.1016/j.proci.2022.06.018}}.

\bibitem{Pavan2022}
C.~A. Pavan, C.~Guerra-Garcia, {Nanosecond pulsed discharge dynamics during
  passage of a transient laminar flame}, Plasma Sources Science and Technology
  31~(11) (2022) 115016.
\newblock \href {https://doi.org/10.1088/1361-6595/aca0bc}
  {\path{doi:10.1088/1361-6595/aca0bc}}.

\bibitem{Pavan2021PlasmaFlames}
C.~Pavan, Y.~Zhang, C.~Guerra-Garcia, {Plasma Actuation of Mesoscale Flames},
  2021 AIAA Aviation Forum, Online (2021) 1--11\href
  {https://doi.org/10.2514/6.2021-3105} {\path{doi:10.2514/6.2021-3105}}.

\bibitem{Pavan2023}
C.~A. Pavan, C.~Guerra-Garcia, {Modeling Flame Speed Modification by Nanosecond
  Pulsed Discharges to Inform Experimental Design}, in: AIAA SCITECH 2023
  Forum, American Institute of Aeronautics and Astronautics, Reston, Virginia,
  2023, pp. 1--15.
\newblock \href {https://doi.org/10.2514/6.2023-2056}
  {\path{doi:10.2514/6.2023-2056}}.

\bibitem{Pavan2022_aviation}
C.~A. Pavan, C.~Guerra-Garcia, D.~Weibel, M.~Nishihara, F.~del Campo,
  S.~Shanbhogue, A.~Ghoniem, {Nanosecond Pulsed Discharge Dynamics in a
  Swirl-Stabilized Combustor with an Unstable Flame}, in: AIAA Aviation 2022
  Forum, American Institute of Aeronautics and Astronautics, Reston, Virginia,
  2022, pp. 1--11.
\newblock \href {https://doi.org/10.2514/6.2022-3574}
  {\path{doi:10.2514/6.2022-3574}}.

\bibitem{Clanet1996}
C.~Clanet, G.~Searby, {On the “tulip flame” phenomenon}, Combustion and
  Flame 105~(1-2) (1996) 225--238.
\newblock \href {https://doi.org/10.1016/0010-2180(95)00195-6}
  {\path{doi:10.1016/0010-2180(95)00195-6}}.

\bibitem{Gonzalez1992}
M.~Gonzalez, R.~Borghi, A.~Saouab, {Interaction of a flame front with its
  self-generated flow in an enclosure: The “tulip flame” phenomenon},
  Combustion and Flame 88~(2) (1992) 201--220.
\newblock \href {https://doi.org/10.1016/0010-2180(92)90052-Q}
  {\path{doi:10.1016/0010-2180(92)90052-Q}}.

\bibitem{Zhong2021_PSST}
H.~Zhong, M.~N. Shneider, X.~Mao, Y.~Ju, {Dynamics and chemical mode analysis
  of plasma thermal-chemical instability}, Plasma Sources Science and
  Technology 30~(3) (2021) 035002.
\newblock \href {https://doi.org/10.1088/1361-6595/abde1c}
  {\path{doi:10.1088/1361-6595/abde1c}}.

\bibitem{Rusterholtz2013UltrafastDischarges}
D.~L. Rusterholtz, D.~A. Lacoste, G.~D. Stancu, D.~Z. Pai, C.~O. Laux,
  {Ultrafast heating and oxygen dissociation in atmospheric pressure air by
  nanosecond repetitively pulsed discharges}, Journal of Physics D: Applied
  Physics 46~(46) (2013) 464010.
\newblock \href {https://doi.org/10.1088/0022-3727/46/46/464010}
  {\path{doi:10.1088/0022-3727/46/46/464010}}.

\bibitem{Xu2014ThermalAir}
D.~A. Xu, M.~N. Shneider, D.~A. Lacoste, C.~O. Laux, {Thermal and hydrodynamic
  effects of nanosecond discharges in atmospheric pressure air}, Journal of
  Physics D: Applied Physics 47~(23) (2014) 235202.
\newblock \href {https://doi.org/10.1088/0022-3727/47/23/235202}
  {\path{doi:10.1088/0022-3727/47/23/235202}}.

\end{thebibliography}





\end{document}